\documentclass[acmtog,nonacm,balance=false]{acmart}
\AtBeginDocument{%
  }

\acmJournal{TOG} % Comment out journal to get conference info in bibstrip.
\usepackage{graphicx}
\usepackage{placeins}
\usepackage{diagbox}
\usepackage{multirow}
\usepackage{makecell}
\usepackage{caption}
\usepackage{subcaption}
\usepackage{algorithm}
\usepackage{algpseudocode}
\usepackage{adjustbox}
\usepackage{wrapfig}
\usepackage{calc}
\usepackage{tabularx}
\usepackage{enumitem}%自定义编号格式

\usepackage{amsmath}       % 数学符号
\usepackage{pifont}
\usepackage{xcolor}        % 颜色支持
\begin{document}

\title{DC-SQEM: Medial Mesh Extraction by Dual Contouring with Spherical Quadric Error Metric}

\author{Bingchuan Li}
\orcid{0009-0004-8355-278X}
\affiliation{%
  \department{School of Informatics}
  \institution{Xiamen University}
  \city{Xiamen}
  \state{Fujian}
  \country{China}
}
\email{libingchuan@stu.xmu.edu.cn}

\author{Ruilong Liu}
\affiliation{%
  \department{School of Film}
  \institution{Xiamen University}
  \city{Xiamen}
  \state{Fujian}
  \country{China}
}
\email{dgmylrl@xmu.edu.cn}

\author{Tianhao Hu}
\affiliation{%
  \department{School of Film}
  \institution{Xiamen University}
  \city{Xiamen}
  \state{Fujian}
  \country{China}
}
\email{hutianhao@stu.xmu.edu.cn}

\author{Congyi Zhang}
\affiliation{%
  \department{Department of Computer Science}
  \institution{The University of Texas at Dallas}
  \city{Richardson}
  \country{USA}
}
\email{congyi.zhang@utdallas.edu}

\author{Ningna Wang}
\affiliation{%
  \department{Department of Computer Science}
  \institution{The University of Texas at Dallas}
  \city{Richardson}
  \country{USA}
}
\email{ningnalala@gmail.com}

\author{Junfeng Yao}
\authornote{Corresponding author.}
\affiliation{%
  \department{School of Film}
  \institution{Xiamen University}
  \city{Xiamen}
  \state{Fujian}
  \country{China}
}
\email{yao0010@xmu.edu.cn}

\author{Xiaohu Guo}
\affiliation{%
  \department{Department of Computer Science}
  \institution{The University of Texas at Dallas}
  \city{Richardson}
  \country{USA}
}
\email{xguo@utdallas.edu}


\begin{abstract}
The \textit{medial axis}, as a compact representation, describes not only the boundary but also the interior of the shape, and has been widely used in analyzing and simplifying shapes. 
However, approximating feature-preserving medial axis from unstructured data, such as oriented point clouds, remains challenging due to the lack of explicit connectivity and the presence of noise in the source.
We propose DC-SQEM, a dual-contouring-based framework that computes medial meshes (an explicit representation of the medial axis) directly from oriented point clouds without relying on watertight manifold surfaces or volumetric structures.
With the help of an adaptive octree, our method fits local medial spheres for each cell based on the Spherical Quadric Error Metric (SQEM).
We address several long-standing robustness issues in dual-contouring and mesh extraction, including out-of-cell solutions, unstable feature classification, and redundant topological connections. To this end, we introduce virtual nodes, a feature-downgrading strategy, and a geometry-driven topology-aware cleanup procedure, enabling robust SQEM-based medial dual contouring with clean mesh outputs.
Unlike prior point-cloud-based methods that produce either only 1D skeletal structures or incomplete medial features, our method preserves sharp geometric features from the input and reconstructs a complete medial structure, including sheets, curves, seams, and junctions.
Compared to restricted-power-diagram-based approaches, our algorithm can handle point clouds as input, without having a strong requirement on water-tightness or manifoldness.
\end{abstract}





\begin{teaserfigure}
    \includegraphics[width=\textwidth]{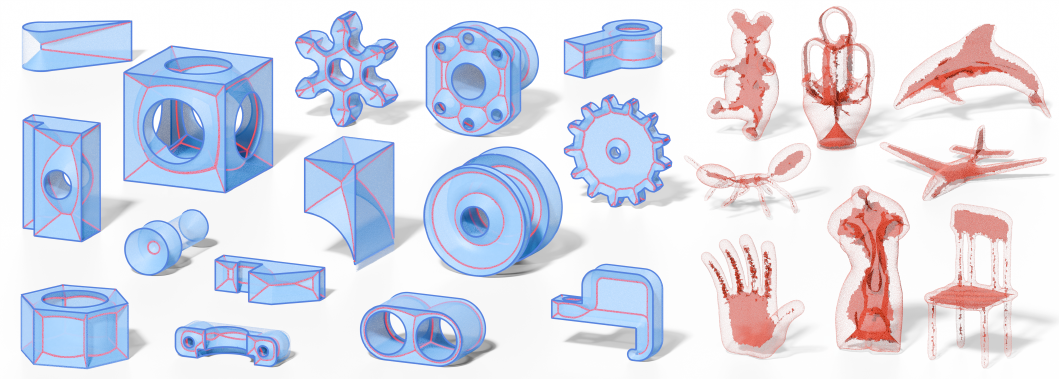}
    \caption{Gallery of extracted medial meshes on a collection of CAD (left) and organic (right) shapes represented as point cloud inputs. We highlight the extracted medial seams as red curves for those CAD models, to demonstrate the structural accuracy of the computed medial surface.}
    \label{fig:gallery}
\end{teaserfigure}

\maketitle

% \begin{figure*}[htbp]
%     \centering
%     \begin{subfigure}[t]{0.23\textwidth}
%         \includegraphics[width=\textwidth]{pics/input.png}
%         \caption{}
%         \label{fig:sub1}
%     \end{subfigure}
%     \hfill
%     \begin{subfigure}[t]{0.23\textwidth}
%         \includegraphics[width=\textwidth]{pics/octree-mesh.png}
%         \caption{}
%         \label{fig:sub2}
%     \end{subfigure}
%     \hfill
%     \begin{subfigure}[t]{0.23\textwidth}
%         \includegraphics[width=\textwidth]{pics/quad-mesh.png}
%         \caption{}
%         \label{fig:sub3}
%     \end{subfigure}
%     \hfill
%     \begin{subfigure}[t]{0.23\textwidth}
%         \includegraphics[width=\textwidth]{pics/tri-mesh.png}
%         \caption{}
%         \label{fig:sub4}
%     \end{subfigure}
%     \caption{xx}
%     \label{fig:four_images}
% \end{figure*}

\section{INTRODUCTION}

The \textit{medial axis} was first introduced by Blum~\cite{blum1967transformation} as a skeleton-like representation of shape defined by the centers of interior maximal balls. Unlike purely boundary-based descriptions, the medial axis simultaneously encodes local thickness, global topology, and salient geometric features. Consequently, it has become a fundamental tool for tasks such as shape editing~\cite{tam2003shape}, recognition~\cite{hu2019matnet},  deformation~\cite{lan2020medial}, segmentation~\cite{lin2020seg}, animation~\cite{yang2018dmat}, and hexahedral meshing~\cite{zhang2024hexmeshing}.
However, computing high-quality Medial Axis Transforms (MATs) from discrete inputs remains a significant challenge. Its extreme sensitivity to noise and intrinsically non-manifold structure make it difficult to balance geometric accuracy with topological reliability, structural clarity, and sharp-feature preservation. For CAD models, sharp edges, corners, and internal structures such as seams and junctions further complicate medial mesh reconstruction.

Existing methods usually make trade-offs among input requirements, topological quality, geometric accuracy, and structural expressiveness. Traditional Voronoi-, scale-axis-, or voxel-based methods~\cite{amenta2001power,miklos2010discrete,yan2018voxel} can approximate the medial axis under certain sampling or resolution conditions, but they may produce redundant sheets, spikes, or topological errors in complex CAD features, thin structures, or local tubular regions. Point-cloud-based or neural methods~\cite{clemot2023neural,huang2024dynamic} can process point cloud inputs, but their outputs are often biased toward 1D skeletons or sparse medial approximations, making it difficult to stably recover complete 2D medial sheets. Recently, Restricted Power Diagram (RPD)-based methods~\cite{wang2022computing,wang2024mattopo,wang2025matstruct} have made significant progress in feature preservation, topology control, and structure optimization for CAD models. However, they typically rely on high-quality watertight surface meshes or tetrahedral volume meshes, which limits their direct applicability to oriented point cloud inputs. A more detailed classification and property comparison is provided in Sec.~\ref{sec:related} and Tab.~\ref{tab:method_comparison_transposed}.

In this paper, we focus on the following problem: how to directly extract geometrically accurate, feature-preserving 3D medial meshes with low topological error rates from oriented point clouds. A high-quality 3D medial representation should not only contain 2D sheets, but also reasonably represent curves, seams, and junctions. Meanwhile, it should preserve sharp external features, avoid excessive cross-sheet connections and redundant volumetric cavities, and remain stable under different sampling densities and complex local geometries.

To this end, we propose \emph{DC-SQEM}, a framework for extracting medial meshes directly from oriented point clouds. 
The key observation behind our method is that the medial axis is the singular structure of the distance field induced by the surface. Away from the medial axis, a spatial point is locally associated with a single closest surface region, and the distance-field gradient varies smoothly. 
Near the medial axis, multiple surface regions compete as closest projections, and the gradient structure becomes locally ambiguous or discontinuous. This suggests that medial-axis extraction can be formulated in a Dual Contouring spirit: instead of detecting sign changes of an implicit function, we build a local spherical quadratic error metric (SQEM)~\cite{thiery2013sphere} from projection constraints, use its singular-value structure to infer whether the local solution behaves as a point, a line, or a plane, and extract representative dual points from valid cells.

However, directly applying the standard Dual Contouring framework to medial-axis extraction is not sufficient. 
First, a valid medial cell does not necessarily provide a stable dual point inside the current cell: the SQEM minimizer may lie outside the cell, and line- or plane-type solution spaces may only intersect neighboring cells. Second, local type classification based only on singular-value thresholding is sensitive to noise and sampling density, which may lead to an overly complex or overly restrictive local type. Third, unlike surface extraction, medial-axis extraction lacks a simple zero-crossing criterion on grid edges, so active cells and their connectivity must be determined from local projection structure rather than from sign changes. 
These differences make it necessary to combine local SQEM fitting with additional geometric and topological mechanisms.

DC-SQEM addresses these issues with three main designs. First, we introduce the concept of \emph{virtual nodes} into the dual contouring process. When an SQEM solution lies outside the current cell but can be consistently represented by a neighboring real node, the current cell can keep a virtual reference to that solution, completing otherwise incomplete four-cell stencils and reducing local gaps. Second, we select candidate non-empty cells using projection clustering on the surface graph, and use the number of projection groups as an upper-bound constraint during SQEM-based type determination. This combines geometric evidence from surface projections with algebraic evidence from singular values, improving the robustness of local node classification. Third, we apply topology-aware and geometry-guided cleanup to the initial quadrilateral medial mesh. By detecting solid-cell complexes, preserving feature-supporting faces, and using a medial-face error metric to guide face removal and triangulation, the method removes redundant cross-sheet or off-medial faces while retaining thin and geometrically faithful medial sheets.

Extensive experiments on the ABC CAD dataset~\cite{koch2019abc} and various organic shapes demonstrate that DC-SQEM performs comparably to state-of-the-art methods across multiple evaluation metrics, including Medial Structural Error Rate (MSER), Topological Error Rate (TER), and Triangle Quality (TQ), despite relying solely on point cloud inputs. Furthermore, DC-SQEM exhibits computational efficiency while maintaining surface reconstruction quality comparable to existing methods. These results establish our framework as a practical and robust tool for high-fidelity medial surface extraction from unstructured point clouds. The technical contributions of this paper are summarized as follows:

\begin{itemize}

\item We propose DC-SQEM, a medial surface extraction framework that operates directly on oriented point clouds. The method combines adaptive octree subdivision, local SQEM fitting, Dual Contouring, and post-processing to generate medial meshes from point-based inputs.

\item We introduce virtual nodes into the dual contouring framework to handle SQEM solutions that lie outside their current cells. These virtual references complete otherwise incomplete four-cell stencils and reduce local connectivity gaps in complex non-manifold medial structures.

\item We propose a non-empty node determination and SQEM-based node type update strategy. By combining surface-graph projection clustering, singular-value analysis, and projection-group constraints, the method improves the robustness of local node classification and reduces missing faces or inaccurate internal feature lines caused by missed non-empty nodes.

\item We propose a topology-aware thinning and geometry-guided cleanup procedure for the initial quadrilateral medial mesh. By combining solid-cell removal, feature-preserving rules, and a medial-face error metric, the method removes redundant cross-sheet or off-medial faces while preserving thinness and improving the geometric quality of the final triangulated medial mesh.
\end{itemize}

\section{RELATED WORKS}
\label{sec:related}

In this section, we review the research most relevant to our proposed framework 
and to the baseline methods used in our experiments. 
We begin with a brief overview of classical medial approximations based on Voronoi diagrams, scale-axis filtering,
and voxel representations, followed by a discussion of Restricted Power Diagram (RPD) methods tailored for feature-preserving and topology-aware medial surfaces
on CAD meshes. Subsequently, we examine approaches that construct medial or skeletal structures directly from point clouds or neural implicit fields.
For a more comprehensive background on the medial axis beyond the scope of this paper, we refer interested readers to the survey by Tagliasacchi et al.~\shortcite{tagliasacchi20163d}.

\subsection{Voronoi- and Voxel-based Medial Approximations}

Early 3D medial axis approximations primarily relied on
Voronoi and power diagrams. A foundational approach, Power Crust~\cite{amenta2001power}, constructs poles and polar balls from the Voronoi diagram of sampled surface points. By utilizing a power diagram for both surface reconstruction and medial approximation, it provides topological and geometric convergence guarantees under strict sampling conditions. In practice, however, when the geometry is complex or the sampling is imperfect, the resulting medial structure may become dense or unstable.

To address the medial axis's inherent sensitivity to noise, the Scale Axis Transform (SAT)~\cite{miklos2010discrete} introduces a scale-dependent skeleton by scaling medial balls and filtering unstable branches.
While SAT is a seminal framework for scale-space representations, its main goal is stable simplification rather than the preservation of clean medial sheet structures or sharp features in CAD models.

In volumetric settings, Voxel Cores~\cite{yan2018voxel} utilize boundary voxels as generators to construct an internal Voronoi structure. This method proves that the resulting core becomes
increasingly thin and homotopy-equivalent to the true medial axis as resolution increases. While Voxel Cores provide strong
theoretical guarantees, they typically require voxelization or closed surface inputs, and high-resolution results may contain dense redundant sheets or spikes. Therefore, VC is more suitable when a reliable voxelization is available, but less direct for our setting of extracting compact, feature-preserving medial meshes from oriented point clouds.

\subsection{RPD-based Feature-preserving Medial Surfaces}

For CAD models characterized by sharp edges and corners, methods based on Restricted Power Diagrams (RPDs) have significantly improved feature preservation and medial mesh quality.
MATFP~\cite{wang2022computing} utilizes a surface RPD to classify medial spheres and defines an energy functional to jointly manage external sharp features and internal medial structures. 
It can preserve medial features and reconstruct sharp CAD features well, but it does not provide a global topological guarantee.
MATTopo~\cite{wang2024mattopo} extends this direction by
introducing a volumetric RPD and local Euler characteristic checks to pursue homotopy equivalence between the medial mesh and the input shape. More recently, MATStruct~\cite{wang2025matstruct} integrates
structure-aware optimization and SQEM-based projections to further enhance sheet clarity, seam/junction consistency,
and triangle quality.
These methods achieve strong results on high-quality mesh inputs, but they require watertight surface meshes or tetrahedral volume meshes and involve complex RPD computations.

Beyond the RPD family, energy- and flow-based methods minimize distance- or curvature-related energies to extract stable internal skeletons~\cite{li2015q,pan2019q,chu2019topology}. However, these are generally optimized for smooth geometries and lack the explicit modeling required for sharp CAD features and strict medial feature preservation.

While RPD-based methods are the current state-of-the-art for manifold mesh or volumetric mesh inputs, their reliance on high-quality watertight surfaces or tetrahedral meshes limits their direct applicability to oriented point clouds.

\subsection{Medial and Skeletal Extraction from Point Clouds}

When only point clouds are available, one research trajectory focuses on extracting skeletal structures directly from these representations. Early methods like the L1-medial skeleton~\cite{huang2013l1} treat skeletonization as a clustering and geometric optimization problem. By minimizing the $L_1$ distance between points and 1D prototypes, these methods robustly recover curve skeletons from noisy data. Such methods are mainly designed for compact curve skeleton extraction, which is different from our goal of constructing explicit two-dimensional medial sheets.

VMAS~\cite{huang2024dynamic} approaches the problem via variational clustering, alternating between Gauss-Newton updates for sphere positions and point-to-sphere assignments. VMAS provides interactive control over the number and distribution of medial spheres, making it suitable for controllable skeleton or medial sphere sampling. In contrast, our method focuses on constructing a sheet-based medial mesh with explicit cell connectivity from oriented point clouds.

Recently, deep learning methods have gained traction. P2MAT-Net~\cite{yang2020p2matnet} learns the pattern of sparse
point clouds and transform them into spheres approximating MAT. Point2Skeleton~\cite{lin2021point2skeleton} and Point2MM~\cite{ge2023point2mm} use neural encoders to predict skeletal points or medial mesh elements and infer their connectivity through graph networks. Neural Skeleton~\cite{clemot2023neural} combines implicit neural representations with coverage formulations to sample skeletal points inside the shape. These methods provide flexible point-cloud or neural-field skeletonization pipelines. Our method instead follows a purely geometric route and explicitly constructs medial sheets through the octree dual complex.

A related direction embeds medial data into neural implicit fields. Deep Medial Fields~\cite{rebain2021deep} and Q-MDF~\cite{kong2026quasi} learn quasi-medial distance fields to recover medial membranes from isosurfaces. Notably, Q-MDF employs a double-covering technique to extract a thin envelope wrapping the medial axis. However, because these membranes are inherently double-layered, extracting a clean
non-manifold 2D mesh remains difficult. Furthermore, these implicit methods offer limited control over sheet manifoldness and medial structure correctness.

Finally, the Coverage Axis family~\cite{dou2022coverage,wang2024coverage} utilizes a set-cover perspective, selecting a sparse subset of interior balls to approximate the medial axis. While these methods provide excellent reconstruction error control, they prioritize coverage over the construction of topologically clean, high-quality medial surface meshes.

In contrast to the aforementioned point-based and neural methods, DC-SQEM is a purely geometric framework driven by local projection clustering and SQEM fitting. It operates directly on oriented point clouds, constructs a dual cell complex on an adaptive octree, and generates sheet-like medial meshes through Dual Contouring and topology-aware cleanup.

\subsection{Property Comparison}
The methods discussed above have different input assumptions and output goals. Some methods primarily target compact curve skeletons, some focus on voxel- or Voronoi-based medial approximations, while RPD-based methods are designed for feature-preserving medial meshes on high-quality mesh inputs. Our DC-SQEM follows a different setting: it directly takes oriented point clouds as input, without relying on watertight surface meshes, tetrahedral meshes, or RPD computation, while aiming to generate explicit medial sheet structures rather than only one-dimensional skeletons. Since our experiments compare representative methods from all these categories, we further summarize them using a unified set of properties in Tab.~\ref{tab:method_comparison_transposed}.

We adopt the properties defined in Tab.~1 of MATTopo~\cite{wang2024mattopo}, namely homotopy preservation, medial feature preservation, thinness, centeredness, reconstructibility, and scalability, while introducing the input type as an additional criterion. We do not repeat their detailed definitions here, but briefly explain the rationale behind the entries. The theoretical properties of PC and SAT generally depend on sampling conditions or scale parameters, so several properties are only conditionally satisfied. VC has stronger theoretical support for homotopy equivalence under sufficiently high voxel resolution, but its scalability is limited by the required voxel resolution. Neural Skeleton and VMAS are flexible point-based methods suitable for skeletonization or medial sphere sampling, but their goal is not to explicitly construct sheet-manifold medial meshes; therefore, they do not fully satisfy properties such as thinness and medial feature preservation in the sense of medial mesh computation. MATFP preserves medial features and has good reconstructibility on surface mesh inputs, but it does not provide a guarantee on homotopy preservation. MATTopo and MATStruct further improve topology and structural quality using volumetric RPDs, but require volumetric mesh inputs and introduce substantial computational cost. DC-SQEM does not provide a guarantee on homotopy preservation. However, through non-empty node determination, local SQEM solving, virtual nodes, Dual Contouring, and topology-aware cleanup, it preserves medial features, thinness, centeredness, and reconstructibility in practice. Since it directly operates on oriented point clouds and avoids voxelization, tetrahedralization, and RPD computation, it also provides good scalability.

\begin{table}[htbp]
\centering
\caption{Comparison of medial axis and skeleton extraction methods. Methods are listed in rows, and properties in columns. Input types: P = Point Cloud, OP = Oriented Point Cloud, SM = Surface Mesh, VM = Volumetric Mesh. Property abbreviations: H = Homotopy Preservation, MF = Medial Feature Preservation, T = Thinness, C = Centeredness, R = Reconstructibility, S = Scalability. The comparison methods are: PC~\cite{amenta2001power}, SAT~\cite{miklos2010discrete}, VC~\cite{yan2018voxel}, Neural Skeleton (NSK)~\cite{clemot2023neural}, VMAS~\cite{huang2024dynamic}, MATFP~\cite{wang2022computing}, MATTopo~\cite{wang2024mattopo}, MATStruct~\cite{wang2025matstruct}, and Ours. \ding{51} means property fulfilled, $\circledcirc$ means conditionally satisfied, - means not satisfied.}
\renewcommand{\arraystretch}{1.2}
\setlength{\tabcolsep}{4pt}
\begin{tabular}{lccccccc}
\toprule
Method & Input & H & MF & T & C & R & S \\
\midrule
PC       & P  & $\circledcirc$ & $\circledcirc$ & \ding{51} & $\circledcirc$ & $\circledcirc$ & \ding{51} \\
SAT      & SM & $\circledcirc$ & $\circledcirc$ & - & $\circledcirc$ & $\circledcirc$ & \ding{51} \\
VC       & SM & $\circledcirc$ & $\circledcirc$ & \ding{51} & $\circledcirc$ & $\circledcirc$ & - \\
NSK      & P  & -       & -       & -           & $\circledcirc$ & $\circledcirc$ & \ding{51} \\
VMAS     & P  & -       & $\circledcirc$ & -           & $\circledcirc$ & $\circledcirc$ & \ding{51} \\
MATFP    & SM & -       & \ding{51} & \ding{51} & \ding{51} & \ding{51} & - \\
MATTopo  & VM & \ding{51} & \ding{51} & \ding{51} & \ding{51} & \ding{51} & - \\
MATStruct& VM & \ding{51} & \ding{51} & \ding{51} & \ding{51} & \ding{51} & - \\
Ours     & OP & -       & \ding{51} & \ding{51} & \ding{51} & \ding{51} & \ding{51} \\
\bottomrule
\end{tabular}
\label{tab:method_comparison_transposed}
\end{table}

As shown in Tab.~\ref{tab:method_comparison_transposed}, RPD-based methods provide strong medial feature preservation, and MATTopo and MATStruct further provide stronger homotopy guarantees; however, they require volumetric mesh inputs and expensive RPD computation. Point-cloud and neural skeleton methods are more flexible in terms of input, but their primary goals are usually different from explicit medial surface construction. DC-SQEM targets the setting between these two families: it avoids volumetric mesh/RPD construction while still producing explicit medial sheets rather than only sparse skeletal curves.

\section{Method}
\begin{figure*}[t]
\centering
\includegraphics[width=\textwidth]{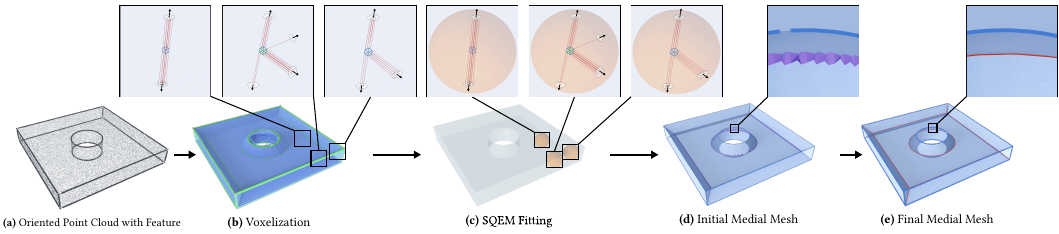}
\caption{\label{fig:pipeline}
Pipeline of DC-SQEM. Given an oriented point cloud with optional feature labels (a), we select candidate octree cells (b), solve SQEM to estimate local medial spheres (c), use virtual nodes and dual contouring to form an initial quad mesh (d), and apply topology- and geometry-aware cleanup to obtain the final triangulated medial mesh (e). 
Non-manifold edges, boundary edges, and solid-cell cavities are shown in \textbf{red}, \textbf{blue}, and \textbf{purple}, respectively.
}
\end{figure*}

Our goal is to reconstruct an explicit medial mesh from an oriented point cloud. For CAD-like inputs where sharp-feature preservation is desired, we optionally assume a feature label for each input point, indicating whether the point lies on a sharp feature; such labels can be obtained by preprocessing methods such as RFEPS~\cite{xu2022rfeps}.

The output of our DC-SQEM is a non-manifold medial mesh $\mathcal{M}$. Each vertex $\mathbf{v}_j=(\mathbf{c}_j, r_j)$ in $\mathcal{M}$ represents a medial sphere with center $\mathbf{c}_j$ and radius $r_j$. They are connected through either edges or triangles to form a non-manifold medial mesh $\mathcal{M}$, which could be further decomposed into either multiple manifold medial \emph{sheets} consisting of $T_2$ spheres or medial \emph{curves} consisting of $T_{\infty}$ spheres, which are connected by medial \emph{seams} of $T_3$ spheres or medial \emph{junctions} of $T_4$ spheres. Here we follow the same notation of $T_2$, $T_3$, $T_4$, $T_{\infty}$ as introduced in MATFP~\cite{wang2022computing}. These decomposed sheets, curves, seams, and junctions represent the structure of the computed medial axis. Similar to MATStruct~\cite{wang2025matstruct}, the medial mesh $\mathcal{M}$ generated by our DC-SQEM contains a clean medial structure.

Fig.~\ref{fig:pipeline} illustrates the overall pipeline of DC-SQEM and highlights its main technical components. 
Given an oriented point cloud with optional feature-point labels (Fig.~\ref{fig:pipeline} (a)), we first construct an octree over the shape domain and retain candidate cells for medial reconstruction (Fig.~\ref{fig:pipeline} (b)). 
On these cells, our SQEM formulation serves as the local fitting primitive: it estimates a medial sphere for each cell, classifies the local solution type, and extracts a representative medial vertex for dual construction (Fig.~\ref{fig:pipeline} (c)). 
To handle cases where the SQEM solution does not yield a point within the current cell, we introduce \emph{virtual nodes} that allow adjacent-cell-supported points to preserve dual connectivity. 
To improve robustness against unstable local type classification, we further combine projection clustering constraints with a feature-downgrading strategy, which prevents overly aggressive point/line/plane assignments from producing invalid local structures. 
The resulting real and virtual nodes are then connected through the grid's dual structure to form an initial quadrilateral medial mesh (Fig.~\ref{fig:pipeline} (d)). 
Finally, because the initial dual mesh may contain redundant connections, solid-cell cavities, and cross-sheet bridges, we apply a geometry-driven, topology-aware cleanup procedure to remove these artifacts while preserving the intended medial topology. 
The cleaned quadrilateral mesh is then triangulated, yielding an explicit medial triangle mesh that preserves sharp features and maintains the non-manifold structure of the medial axis (Fig.~\ref{fig:pipeline} (e)).

In this section, we will introduce the SQEM formulation~\cite{thiery2013sphere} and the virtual-node extension for handling out-of-cell solutions (Sec.~\ref{subsec:sqem}); the candidate-cell selection stage, which avoids unnecessary SQEM solves (Sec.~\ref{subsec:candidate}); the type-downgrading strategy for improving the robustness of local SQEM classification (Sec.~\ref{subsec:downgrading}); and the topology- and geometry-aware cleanup procedure that converts the initial dual mesh into a clean medial mesh (Sec.~\ref{subsec:dualmesh}).

\subsection{Dual Contouring}
Dual Contouring~\cite{ju2002dual} is a cell-centered method for extracting an implicit surface from a spatial grid, typically an octree. Unlike Marching Cubes, which places vertices on grid edges, Dual Contouring constructs a dual representation of the grid cells. Each active cell, usually identified by a sign change of the implicit function on its boundary, contributes one representative vertex. This vertex is commonly computed by solving a local quadratic error function (QEF) from Hermite data, such as intersection points and normals. The final mesh is then assembled by connecting the representative vertices of neighboring active cells around grid edges. This one-vertex-per-cell formulation naturally supports adaptive octrees and can better preserve sharp features through the local QEF solve.

\subsection{Local SQEM}
\label{subsec:sqem}
In this section, we introduce our SQEM-based dual contouring formulation in detail.
For a candidate cell $C$ 
%\ningna{how about using `active' cell vs `non-active' cell here? The `active' cell is the cell that has intersection with underlying implicit medial axis.}\congyi{i think it is too early to introduce those concepts. here we introduce the idea of sqem, and the active/non-active can be defined after solving sqem or after we introduce the global octree pruning.}, 
we aim to estimate its dual vertex -- a representative medial sphere $\mathbf{s}=(\mathbf{c},r)$ inside $C$.

Given a set of sample points $\{\mathbf{x}_i\}_{i=1}^m$ in $C$, we \emph{project} them onto the input point cloud. That is, for any point $\mathbf{x}_i$ in 3D space, we find its closest surface sample $\mathbf{p}_i=\pi(\mathbf{x}_i)$, which could be easily implemented through a kd-tree data structure. For a non-feature surface sample $\mathbf{p}_i$, we define the vector $\mathbf{n}_i$ as its surface normal. While for a feature surface sample, the vector $\mathbf{n}_i$ is computed as the normalized direction from $\mathbf{x}_i$ to $\mathbf{p}_i$. Each pair $(\mathbf{p}_i, \mathbf{n}_i)$ contributes a tangent-plane constraint for the estimation of medial sphere $\mathbf{s}$: 
%\ningna{TODO: how the projection and normal $n$ is made?}
$$
\mathbf{n}_i^{\top} (\mathbf{p}_i - \mathbf{c}) - r = 0.
$$
We compute $\mathbf{s}$ by minimizing the quadric error:
\begin{equation}
  \mathbf{s}^{*} = \arg\min_{(\mathbf{c},r)} \sum_{i=1}^{m} \bigl(\mathbf{n}_i^{\top}  (\mathbf{p}_i - \mathbf{c}) - r\bigr)^2,
  \label{eq:sqem_ls}
\end{equation}
which can be formulated as a standard linear least-squares problem $\min_\mathbf{x} \|\mathbf{A}\mathbf{x} - \mathbf{b}\|^2$ with unknown $\mathbf{x}=(\mathbf{c}^\top,r)^\top$. The matrix $\mathbf{A}$ and vector $\mathbf{b}$ are directly assembled from $\{(\mathbf{p}_i, \mathbf{n}_i)\}_{i=1}^{m}$. We compute the Singular Value Decomposition (SVD) of $\mathbf{A}$ to obtain its singular value spectrum, which could be used to determine the type of medial sphere for cell $C$ and the corresponding constrained geometry.

Our formulation Eq. \ref{eq:sqem_ls} could be extended to other surface input representations such as polygon soups or implicit functions, as long as the projection operation $\pi(\mathbf{x})$ could be robustly implemented. 

As in QEF-based dual contouring, the singular-value structure of this local system indicates whether the solution behaves as an isolated point, a line, or a plane. This provides the type of local primitive for cell $C$ and gives a principled way to extract one dual vertex from the corresponding type.

Let the singular values of the SQEM system in cell $C$ be
\begin{equation}
\sigma_0 \ge \sigma_1 \ge \sigma_2 \ge \sigma_3 \ge 0.
\end{equation}
We call $\sigma_i$ degenerate if $\frac{\sigma_i}{\sigma_0} \le \epsilon$,
where $\epsilon$ is a small threshold.
As discussed above, this threshold alone is not sufficiently robust for singular-value (or cell type) classification. 
We therefore choose a threshold that tends to retain more non-zero singular directions rather than prematurely treating them as degenerate. In our implementation, we set $\epsilon = 10^{-2}$.
In the following stages, the number of projection clusters is used as an additional upper-bound constraint, and a type-downgrading strategy is introduced to further improve robustness, as described in Sec.~\ref{subsec:candidate}.

\begin{figure}
\centering
\includegraphics[width=\linewidth]{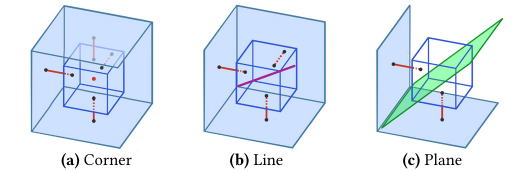}
\caption{\label{fig:sqem_type}
Illustration of SQEM solution types determined by singular-value degeneracy.
(a) Corner type (Case 1): no degenerate singular direction yields an isolated SQEM minimizer.
(b) Edge type (Case 2): one degenerate singular direction yields an approximate line solution space, and the candidate dual point is chosen from the line-cell intersection.
(c) Plane type (Case 3): two degenerate singular directions yield an approximate planar solution space, and the candidate dual point is chosen from the intersection between the plane and the cell.
}
\end{figure}
The number of degenerate singular values and the corresponding singular directions determine the type of the current cell $C$:

\paragraph{Case 1: Corner Type: no degenerate singular direction.}
In this case, the SQEM's solution is an isolated point (see Fig.~\ref{fig:sqem_type}(c)), which corresponds to a $T_4$-type medial sphere, and the solution space is an isolated point. If the center $\mathbf{c}^{*}$ lies inside the current cell $C$, we tentatively mark $C$'s dual as a \textbf{real node}. Otherwise, if the point is inside the face- or edge- adjacent cell (its neighbor cell), we treat $C$'s dual as a \textbf{candidate virtual node}. If $\mathbf{c}^{*}$ does not fall inside $C$ or any of its neighbors, $C$'s dual is treated as an \textbf{empty node}.

\paragraph{Case 2: Line Type: one degenerate singular direction ($\sigma_3 \approx 0$).}
The solution space is approximately constrained to a one-dimensional line $\ell$ (see Fig.~\ref{fig:sqem_type}(b)), corresponding to a $T_3$- or $T_{\infty}$-type sphere lying on a medial seam or a medial curve. We obtain the singular vector associated with $\sigma_3$ and take the least-squares solution $\mathbf{c}^*$ as a point on the line $\ell$. We intersect $\ell$ with the 6 faces of cell $C$. If they intersect, we take the center of the intersected segment as a candidate center $\mathbf{c}^*$ and tentatively mark $C$'s dual as a real node. If $\ell$ does not intersect $C$ but intersects some of its neighboring cells, we record the intersection centers in those neighboring cells, and treat $C$'s dual as a candidate virtual node. If $\ell$ intersects neither $C$ nor any of its neighbors, $C$'s dual is treated as an empty node.

\paragraph{Case 3: Plane Type: two degenerate singular directions ($\sigma_2,\sigma_3 \approx 0$).}
The solution space is approximately a plane $\Pi$ that corresponds to a $T_2$-type sphere lying on a medial sheet. We derive the plane normal $\mathbf{n}$ from the associated singular vectors and enforce $\mathbf{c}^*$ to lie on the plane $\Pi$ (see Fig.~\ref{fig:sqem_type}(a)). We intersect $\Pi$ with the current cell $C$. If $\Pi$ intersects $C$, we compute the intersected polygon, take its centroid as a candidate center $\mathbf{c}^*$, and tentatively mark $C$'s dual as a real node. If $\Pi$ does not intersect $C$, but only intersects some of its neighboring cells, we record the intersected centers in those neighboring cells, and mark $C$'s dual as a candidate virtual node. If there is no intersection with $C$ or any of its neighbors, $C$'s dual is treated as an empty node.

\begin{wrapfigure}{r}{0.5\linewidth}
\centering
\includegraphics[width=1.0\linewidth]{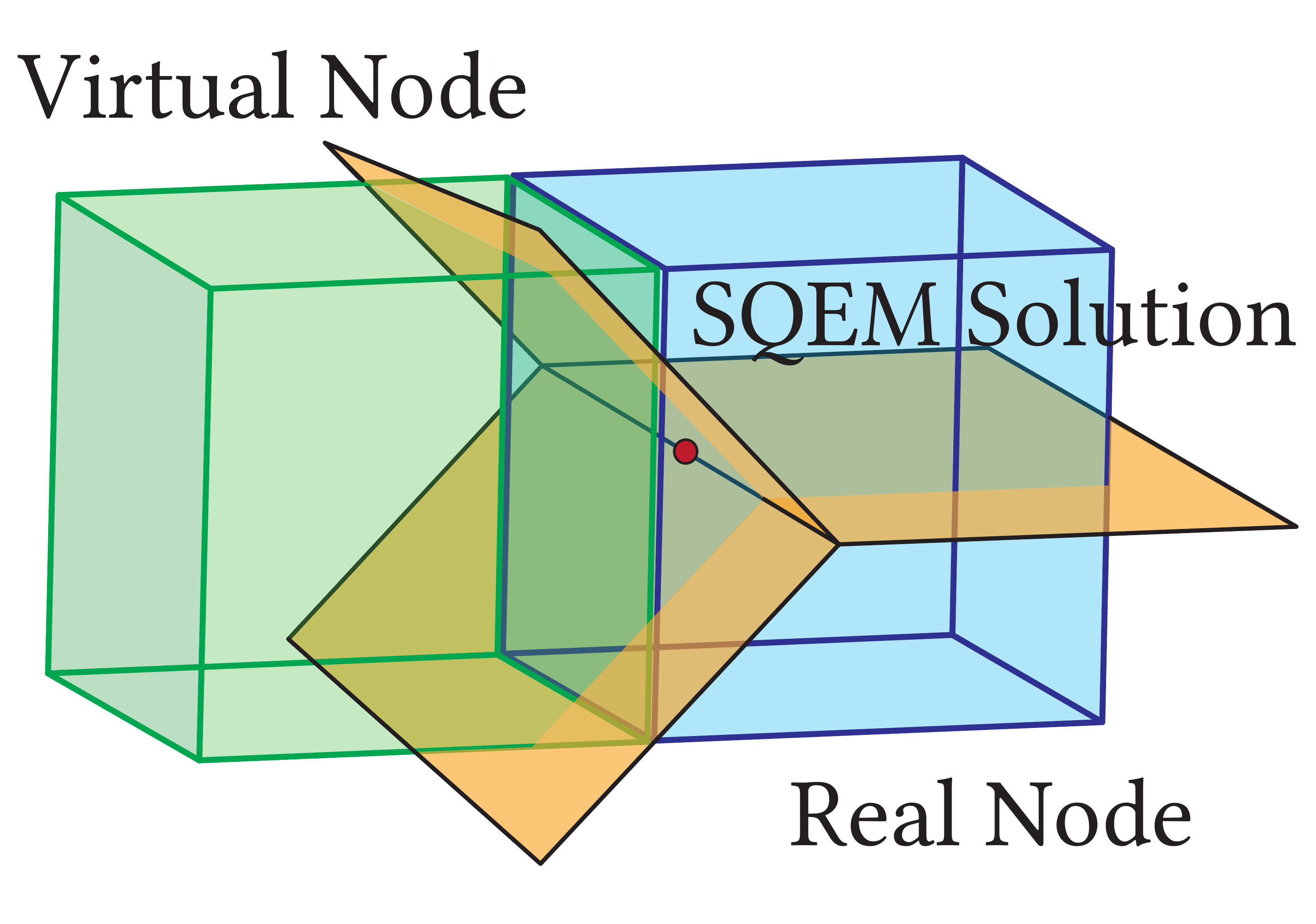}
\caption{\label{fig:virtual_node}Illustration of a virtual node and its referred neighboring real node.}
\end{wrapfigure}

\paragraph{Case 4: Null Type: three degenerate singular directions ($\sigma_1,\sigma_2,\sigma_3 \approx 0$).}
In this extreme degenerate case, the seed sphere is far from the medial axis, and the cell provides almost no constraint on the medial axis position. We directly mark $C$'s dual as an empty node.

Note that, unlike standard dual-contouring-based mesh extraction frameworks~\cite{ju2002dual}, we introduce the concept of a \emph{virtual node} (see Fig.~\ref{fig:virtual_node}) to handle dual points whose SQEM solutions lie outside their current cells. 

This virtual node mechanism helps reduce hole artifacts caused by discarded outside-cell solutions or local SQEM solver instability. In Sec.~\ref{sec:discussion_virtualnodes}, we discuss in depth the situations where the virtual node mechanism fails, as well as why it nevertheless performs well in practice when computing the medial axis for the majority of 3D shapes.

\subsection{Candidate-cell Selection}
\label{subsec:candidate}

With the per-cell computations established, we now outline how candidate cells are selected to avoid unnecessary SQEM evaluations.

We build an adaptive octree over the bounding box of the input point cloud and increase resolution only in \textbf{potential medial regions}. We recursively subdivide cells from coarse to fine: a cell is not further subdivided if it is classified as outside the shape, ``obviously-empty'', or has already reached the maximum depth; after subdivision, only non-empty leaf cells are used for SQEM-based medial sphere fitting.

\paragraph{Projection grouping.} Unlike traditional criteria that rely on whether a zero level set intersects the cell, we use \textbf{clustering of local sample projections} to decide whether a cell may contain the medial axis. Given an octree cell $C$, we uniformly sample a set of points $\{\mathbf{x}_i\}_{i=1}^m$ inside it, project them onto the input point cloud: $\{\mathbf{p}_i\}_{i=1}^m$. Intuitively, if $C$ does not intersect with the medial axis, the projected points $\{\mathbf{p}_i\}$ tend to form a single cluster; when the medial axis passes through $C$, the projected points $\{\mathbf{p}_i\}$ typically form two or more clusters on the input surface. We build a nearest neighboring graph for the points on the input surface, and use a simple connected-component test for the clustering purpose. It is easy to note that such a simple clustering approach is brittle: one cluster may be unexpectedly split into two or more by sampling noise.
%To stabilize the test, we expand each projected sample by one ring based on the nearest neighboring graph, but stop the expansion at sharp features.
%[do we explicitly detect those sharp features?].
However, it is worth noting that the set of candidate cells is intentionally chosen to be slightly conservative: some cells that are only tangentially touched by the medial axis are also retained and later handled by the SQEM fitting and dual contour meshing steps. This design ensures that, even at a relatively-coarse octree resolutions, we do not break cross-cell medial structures due to overly-aggressive empty-cell classification. 

\subsection{Cell Type Downgrading}
\label{subsec:downgrading}

Now we have two sources of cell types: one coming from the singular values of SQEM -- Corner ($T_4$), Line ($T_3$ or $T_{\infty}$), Plane ($T_2$), and Null (empty cell); another one coming from the projection grouping -- four clusters ($T_4$), three clusters ($T_3$), two clusters ($T_2$), and one cluster (empty cell). In this subsection, we discuss how the conservative grouping result mentioned in the above Sec.~\ref{subsec:candidate} could be used to constrain the SQEM classification, thereby improving the robustness of cell type determination.

In practice, if a cell contains four or more projection groups, then a corner-type solution is allowed.
If it contains three groups, then we restrict the solution to a simpler type, such as edge or plane. If only two groups, then we restrict the solution to the plane type only. We allow the cell types initially determined by projection grouping to be \emph{downgradable}, in the order of Corner $\rightarrow$ Line $\rightarrow$ Plane, based on the SQEM solution space.

In this way, the projection grouping serves two purposes: it avoids unnecessary SQEM solves on those empty cells, and it prevents the cell type computed by SQEM from being more complex than what the projection grouping supports.

A candidate cell may pass the grouping test and still fail to accept a dual vertex through SQEM solution (Sec.~\ref{subsec:sqem}).
This happens precisely when the dual vertex lies far outside the cell.
In the medial setting, such cases are common enough that they must be treated as part of the method rather than as numerical exceptions.
We therefore place the local SQEM computation inside an outer loop.
For a non-empty cell $C$, we first determine the cell type under the projection grouping in Sec.~\ref{subsec:candidate}, and then compute the corresponding dual vertex through SQEM as in Sec.~\ref{subsec:sqem}.

If the cell type computed by SQEM fails on the current cell (the cell is neither marked as a real or virtual node), we downgrade the type in the order of Corner $\rightarrow$ Line $\rightarrow$ Plane, and repeat the dual vertex solution with SQEM. If none of these types yields an accepted dual vertex, the cell is finally labeled as empty.

This outer loop is a key part of the method.
It makes the one-vertex-per-cell representation reliable in the medial setting:
a valid cell is not rejected simply because its SQEM-based cell type does not provide a legal dual vertex.
At the same time, the downgrading rule avoids forcing an overly restrictive local type when a simpler type already explains the cell. A detailed algorithm about the type downgrading process is provided in the Supplementary Document.

\subsection{Dual Mesh Construction and Cleanup}
\label{subsec:dualmesh}

\subsubsection{Dual mesh construction}

After the outer loop, each maximum-depth cell is either active or empty. Similar to DualMesh-UDF~\cite{zhang2023surface}, the mesh connection rule is designed as follows. Each edge in the grid is shared by four adjacent cells. If all four adjacent cells are non-empty, we generate one quad connecting the dual points (the medial sphere centers obtained from the solution in Sec.~\ref{subsec:sqem}) of these four cells.

This step is where the above-mentioned design choices come together.
The one-vertex-per-cell representation keeps the dual construction simple, while virtual nodes prevent valid local configurations from leaving holes in the mesh merely because one cell did not provide its own inside-cell dual vertex.

\subsubsection{Topology- and geometry-aware cleanup}
\label{subsubsec:quad-error}
The initial dual mesh captures the main medial structure, but it is still not the final result.
Because the dual mesh is built from dual vertices and local connectivity, it may contain artifacts that are topologically or geometrically inconsistent with the intended medial axis.
We remove them by a topology-aware, geometry-guided cleanup procedure.

\paragraph{Topology-preserving solid-cell deletion.} A \emph{solid-cell complex} is a $2\times2\times2$ block of adjacent active cells.
Such a block creates a volumetric region in the dual mesh due to the edge-based quad mesh construction. The solid-cell complex should not be retained in the final medial mesh. We perform the cleanup of solid-cell complexes by removing \emph{simple pairs} of (solid, face) or (face, edge), inspired by the \emph{simple removal} process of Ju et al.~\shortcite{ju2007editing}. We first break every such complex by deleting at least one of its boundary faces together with the solid cell, i.e., the (solid, face)-simple pair.
After all solid-cell complexes are broken, we continue deleting remaining pocket faces by removing those (face, edge)-simple pairs.
To monitor topology, we use the Euler characteristic:
\begin{equation}
\chi = V - E + F - C,
\end{equation}
where $V$, $E$, $F$, and $C$ are the numbers of vertices, edges, quad faces, and intact solid-cell complexes, respectively.
The simple-pair-removal process guarantees to keep $\chi$ unchanged~\cite{ju2007editing}.

\begin{figure}
\centering
\includegraphics[width=\linewidth]{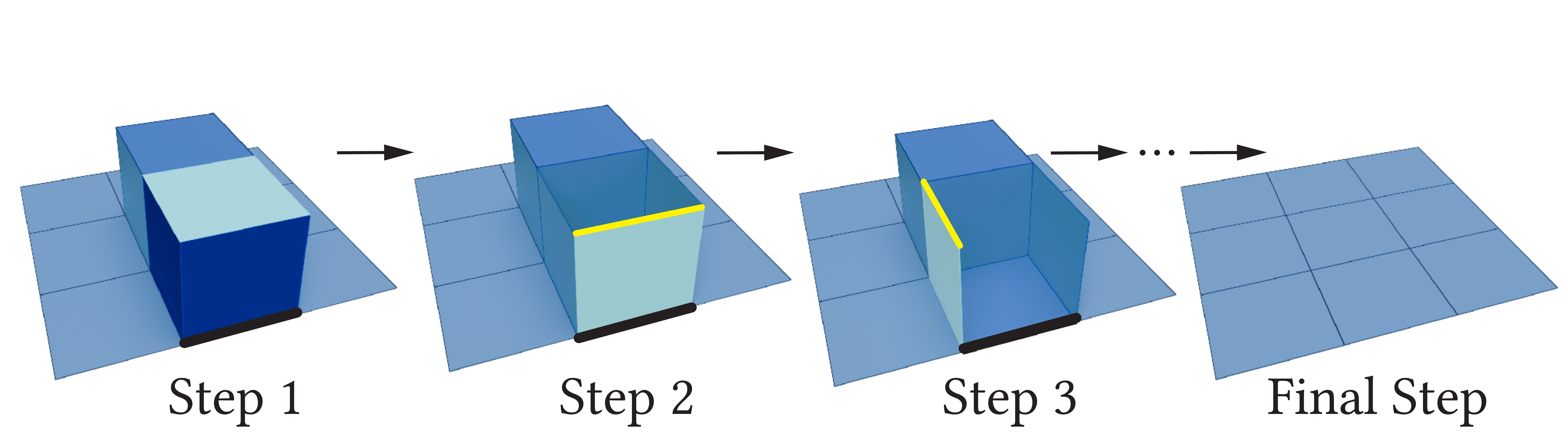}
\caption{\label{fig:thinning}Steps of solid-cell thinning by removing simple pairs, in sequential order. Faces incident to a feature edge (black line) may be simplified, but at least one supporting face per feature edge is always preserved.}
\end{figure}

\paragraph{Feature preservation}
Adopting the terminology from MATFP~\cite{wang2022computing}, we define the \textbf{external feature edges} of the medial mesh as those corresponding to convex sharp features on the input surface, and the \textbf{internal feature edges} as the medial seams connecting distinct medial sheets.
A quadrilateral face is designated as the sole supporting medial face for an external feature edge if it is the only face incident to a boundary edge with feature-marked endpoints. We forbid the removal of such faces to prevent large interior gaps along the feature lines. However, if multiple faces are incident to the same feature edge, we allow their gradual deletion governed by geometric error, subject to the strict constraint that at least one supporting face is preserved for every feature edge. A simple example is shown in Fig.~\ref{fig:thinning}.

\begin{figure}
\centering
\includegraphics[width=\linewidth]{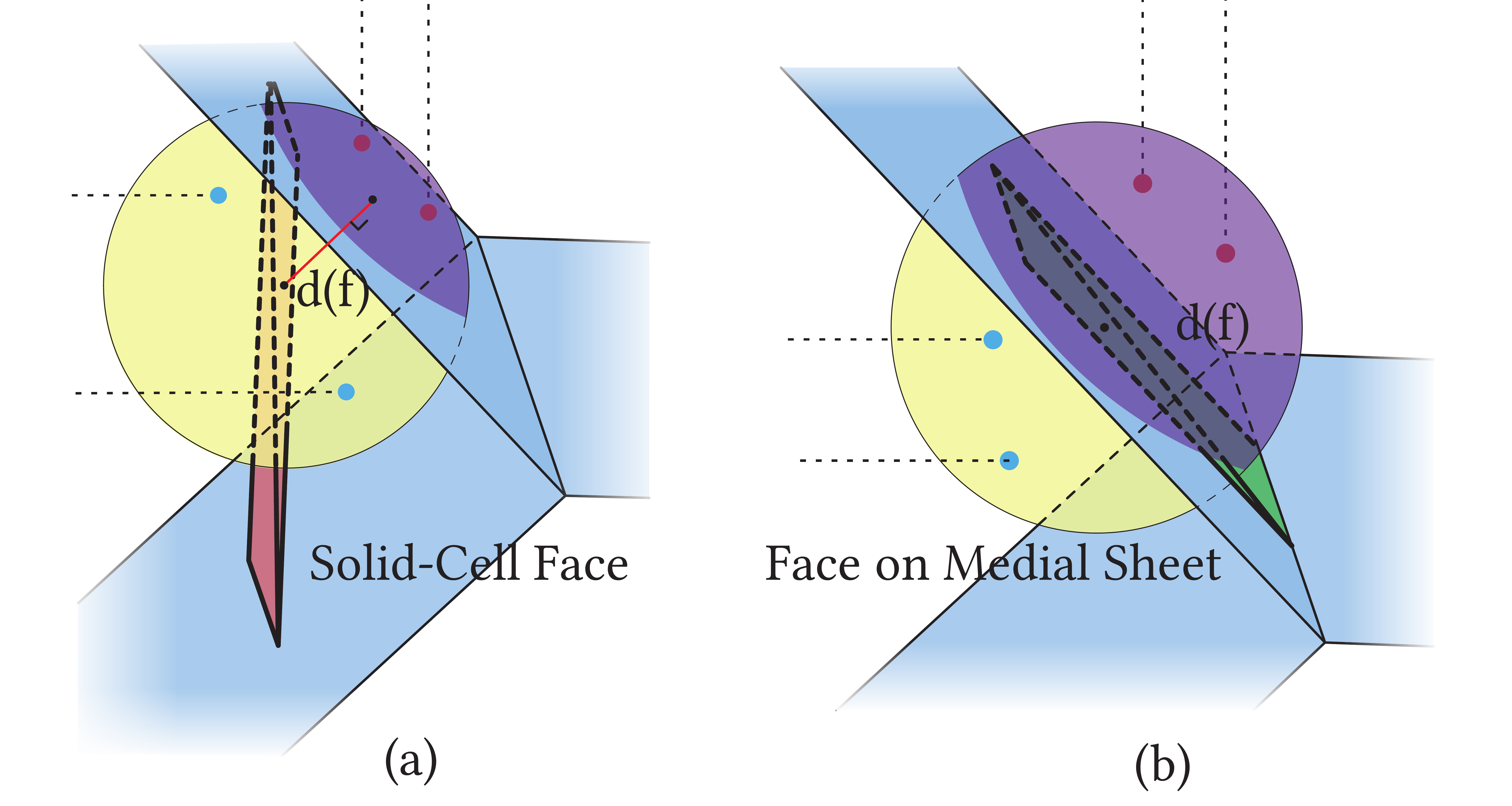}
\caption{\label{fig:face_error}(a): the cross-sheet face of a solid-cell will have a high d(f). (b): the in-sheet face that aligns well with the medial axis will have a low d(f).}
\end{figure}

\paragraph{Geometry error}
We estimate a geometry error $d(f)$ for each quadrilateral face $f$ in the initial medial mesh. 
This error is used in two places: during solid-cell face removal, it defines the geometric priority for removing pocket faces; after cleanup, it provides correct triangulation of the remaining quad faces.
Intuitively, faces with large $d(f)$ are less compatible with a true medial sheet and are therefore preferred for removal when they are topologically removable.

For a quad face $f$, we define
$$
d(f) = \min_{T\in \mathcal{T}(f)} \; \max_{t \in T} dist(t),
$$
where $\mathcal{T}(f)$ denotes the two possible triangulations for the chosen quad, and $T$ denotes the set of two triangles in one potential triangulation. For a triangle $t$, let $c_t$ be its centroid.
We place a small sphere with radius $\epsilon$ centered at $c_t$, sample points inside the sphere, and project them back to the input surface.
If all projected samples belong to one connected projection group, then this triangle is locally supported by only one surface patch and therefore cannot represent a valid medial face.
In that case, we set $\mathrm{dist}(t)=+\infty$.
Otherwise, we fit a local tangent plane $\Pi_t$ from the two largest projection groups and define $\mathrm{dist}(t)$ to be the distance from the centroid $c_t$ to the fitted plane $\Pi_t$:
\begin{equation}
\mathrm{dist}(t)=\mathrm{dist}(c_t,\Pi_t).
\end{equation}
%If $t$ degenerates numerically to a point, we set $\mathrm{dis}(t)=0$.

This design ensures that we only remove solid-cell faces with large errors, most of which are medial faces spanning multiple sheets (see Fig.~\ref{fig:face_error} (a)), while preserving faces that closely approximate the true medial axis (see Fig.~\ref{fig:face_error} (b)). 

\paragraph{Distance-minimizing triangulation.}
For each remaining quadrilateral $f$, we consider its two possible triangulations and use the triangle distance $dist(t)$ to choose between them: for each possible triangulation of $f$, we take the maximum error over its two triangles, and select the triangulation with the smaller maximum error. In this way, the triangulation naturally favors configurations that are closer to the true medial sheets and improves the geometric accuracy of the triangle medial mesh.

% \begin{wrapfigure}{r}{0.5\linewidth}
% \centering
% \includegraphics[width=1.0\linewidth]{picsnew/cluster.pdf}
% \caption{\label{fig:clustering}Illustration of clustering.}
% \end{wrapfigure}

\subsubsection{$T_{\infty}$-type connectivity}
% After cleanup, we remove small isolated patches and degenerate boundary fragments.
% Finally, if an edge-type active cell is not connected through valid faces, we explicitly insert a medial edge to complete the corresponding one-dimensional branch.
% The final output is therefore a unified medial mesh containing both cleaned medial sheets and completed medial curves.

Finally, for detected $T_{\infty}$-type active cells that are not connected through valid faces, we explicitly insert medial edges to improve the connectivity of the corresponding one-dimensional branches.
To detect a $T_\infty$-type node, we first estimate a local axis that approximates the tubular direction of the cell. This axis is computed as the perpendicular direction of the plane spanned by the projection directions of the cell. Once the axis is estimated, we evaluate the angular coverage of the projection directions around this axis to determine whether the cell corresponds to a thin and elongated medial curve. For complete algorithmic details for the detection of $T_{\infty}$-type cells, including clustering, projection, thresholding, and fallback strategies, we refer readers to the Supplementary Document.
The final output is therefore a unified medial mesh containing  triangulated medial sheets and locally completed medial curves.

\section{Experiments}
\label{sec:experiments}

We present quantitative and qualitative evaluations of the proposed method. Our evaluation dataset consists of 93 CAD models from the ABC dataset~\cite{koch2019abc} alongside 8 organic models. We deliberately adopt the exact same subset of 93 CAD models used in MATStruct~\cite{wang2025matstruct} to maintain consistency with prior literature and prevent any dataset selection bias. Prior to processing, all input shapes are normalized to fit inside the $[-0.5,0.5]^3$ bounding box. The proposed algorithm is implemented by calling C++ modules from Python, and all baselines are run using the authors' original code. The source code for DC-SQEM is available at \url{https://github.com/Jiafeimaochuanqi/dc_sqem.git}. All experiments are performed on a workstation with an Intel(R) Core(TM) i7-14700KF CPU at 3.40 GHz and 64 GB RAM. 

\begin{table}[htbp]
    \centering
    \caption{Quantitative comparison of different baselines on 93 CAD models extracted from ABC dataset. Best values are shown in \textbf{bold}, second-best values are \underline{underlined}, and third-best values are shown in \textit{italics}.}
    \label{tab:metrics_portrait}
    \setlength{\tabcolsep}{5pt}
    \renewcommand{\arraystretch}{1.25}

    \begin{tabular}{l|l|l|l|l|l}
        \hline
        \footnotesize\textbf{Method} 
        &\footnotesize\textbf{MSER(\%)$\downarrow$} 
        &\footnotesize\textbf{TQ$\uparrow$} 
        &\footnotesize\textbf{TER(\%)$\downarrow$} 
        &\footnotesize\textbf{HD(\%)$\downarrow$} 
        &\footnotesize\textbf{SITR(\%)$\downarrow$}\\
        \hline

        \textbf{Ours}&\underline{3.0874} & \textit{0.683} &\underline{2.1505}&\textbf{0.4203}&4.0711 \\
        \hline

        PC\_160&94.0065&0.486&100&1.1668&\textbf{0}\\
        PC\_200&94.2978&0.487&100&\textit{1.1219}&\textbf{0}\\
        NSK&98.3696&0.653&100&--&\textbf{0}\\
        SAT-0.03&84.0117&0.269&100&4.4955&20.3429\\
        SAT-0.04&85.0257&0.254&100&4.7628&16.8332\\
        VMAS&95.6633&\underline{0.695}&100&1.9718&10.4908\\
        VC&47.2684&0.599&\textit{3.0928}&1.7512&\underline{0.0009}\\
        MATFP&\textit{32.8499}&0.533&37.5&1.3195&\textbf{0}\\
        MATTopo&66.8900&0.485&\textbf{0}&1.2545&\textbf{0}\\
        MATStruct&\textbf{0.9423}&\textbf{0.712}&\textbf{0}&\underline{1.0918}&\textit{\textit{0.0064}}\\
        \hline
    \end{tabular}

    \vspace{0.8ex}
    {\small Notice: -- indicates missing values returned by the skeleton-based method.}
\end{table}

\subsection{Baselines}
\label{subsec:baselines}

We compare our method with eight state-of-the-art (SOTA) approaches: PC~\cite{amenta2001power}, SAT~\cite{miklos2010discrete}, VC~\cite{yan2018voxel}, Neural Skeleton (NSK)~\cite{clemot2023neural}, VMAS \cite{huang2024dynamic}, MATFP~\cite{wang2022computing}, MATTopo \cite{wang2024mattopo}, and MATStruct~\cite{wang2025matstruct}. Among them, Neural Skeleton and VMAS are skeletonization methods that primarily aim to produce sparse medial representations (typically with only a few hundred balls), while the others are medial-axis approximation methods. 

We include VMAS in our comparison because its input is also point clouds. However, when the target number of balls increases to several thousands, VMAS becomes unstable: the optimization exhibits oscillatory behavior, repeatedly inserting and deleting balls without converging. For a fair comparison, we run VMAS on dense surface points (each with more than 600K points) and gradually increase the target number of spheres. If the number of spheres does not change for 100 consecutive iterations, we terminate the optimization and use this configuration as the maximal stable capacity of VMAS for evaluation.

\subsection{Evaluation Metrics}
\label{subsec:metrics}

To evaluate the quality of the generated medial meshes, we adopt four metrics: \textbf{medial structural error rate (MSER)}~\cite{wang2025matstruct}, \textbf{topological error rate (TER)}~\cite{wang2025matstruct}, \textbf{self-intersecting triangle ratio (SITR)}, and \textbf{triangle quality (TQ)}~\cite{frey1999surface}. We also use the \textbf{Hausdorff distance (HD)} to assess the geometric fidelity between reconstructed surfaces and the original inputs. 
%Specifically, we use $\epsilon^1$ to denote the one-sided Hausdorff distance from the original surface to the MAT-reconstructed surface, and use $\epsilon^2$ for the distance in reverse side.

\paragraph{Medial Structure Error Ratio (MSER)}
MSER~\cite{wang2025matstruct} quantifies the percentage of misclassified \textbf{seam} and \textbf{junction} spheres on medial mesh. Fundamentally, a regular medial sphere on a sheet contacts the surface at exactly two points, while seam and junction spheres must possess three or more distinct tangential points to represent structural transitions. We employ a robust iterative search strategy to recover these contact points. A detected seam or junction sphere is considered \textbf{misclassified} if the number of its distinct tangential points does not exceed that of its neighboring sheet spheres.
MSER is defined as the ratio of these misclassified singularities to the total number of seam and junction spheres. We restrict this evaluation to CAD models, where such medial structures like seam and junctions are unambiguously defined.

\paragraph{Topology Error Ratio (TER)}
TER~\cite{wang2025matstruct} denotes the percentage of medial meshes that fail to preserve the ground-truth topology. A medial mesh is considered \emph{topologically correct} if and only if both the number of connected components (i.e., the 0th Betti number $b_0$) and the Euler characteristic $\chi$ match those of the ground-truth. Given that there is no cavities in any of our dataset's 3D shapes (i.e., the 2nd Betti number $b_2 = 0$), achieving consistency in both $b_0$ and $\chi$ inherently ensures consistency in  the 1st Betti number $b_1$, according to the Euler-Poincar\'e formula $\chi=b_0-b_1+b_2$. TER is then defined as the fraction of computed medial meshes that do not satisfy the topological correctness.

\paragraph{Triangle Quality (TQ)}
TQ~\cite{frey1999surface} evaluates mesh regularity by quantifying the deviation of each triangle from a perfect equilateral one, where higher values indicate better element quality.

\paragraph{Hausdorff Distance (HD)}
HD evaluates the reconstruction error of the computed medial mesh, as a percentage relative to the diagonal length of the input shape's bounding box. Specifically, we report two one-sided Hausdorff distances. $\epsilon^1$ represents the distance from the ground truth surface to the MAT-reconstructed surface, while $\epsilon^2$ represents the distance in reverse side.

\paragraph{Self-Intersecting Triangle Ratio (SITR)}
SITR measures the percentage of triangles involved in at least one self-intersection with a non-adjacent triangle. We detect all intersecting triangle pairs, discard pairs sharing mesh vertices, and collect the unique triangle indices appearing in the remaining pairs. SITR is defined as the number of such self-intersecting triangles divided by the total number of triangles in the medial mesh. Lower values indicate fewer geometric self-intersections.

\begin{figure}
\centering
\includegraphics[width=\linewidth]{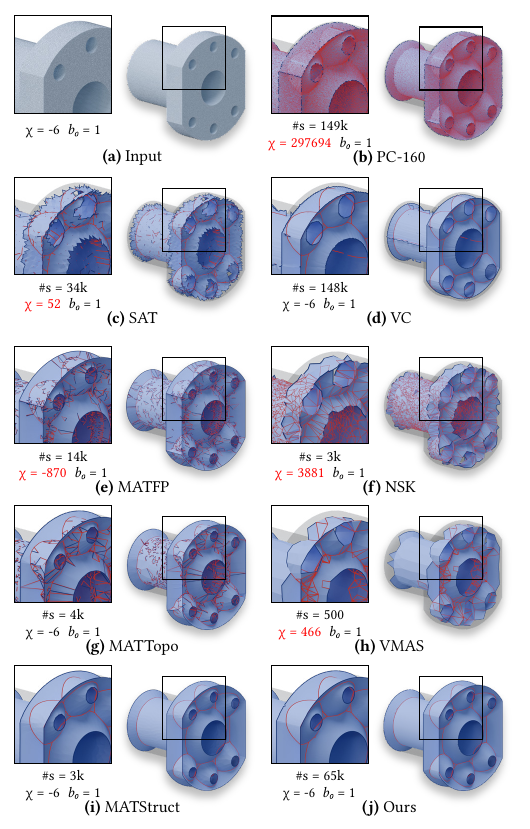}
\caption{Visual comparison of topological fidelity on a sharp corner shape. We report the Euler characteristic ($\chi$) and the number of medial balls ($\#s$) for each method. Incorrect topological values are marked in \textbf{\textcolor{red}{red}}. Non-manifold edges in the medial meshes are highlighted as \textbf{\textcolor{red}{red}} lines.}
\label{fig:topo_comparison}
\end{figure}

\begin{figure}
\centering
\includegraphics[width=\linewidth]{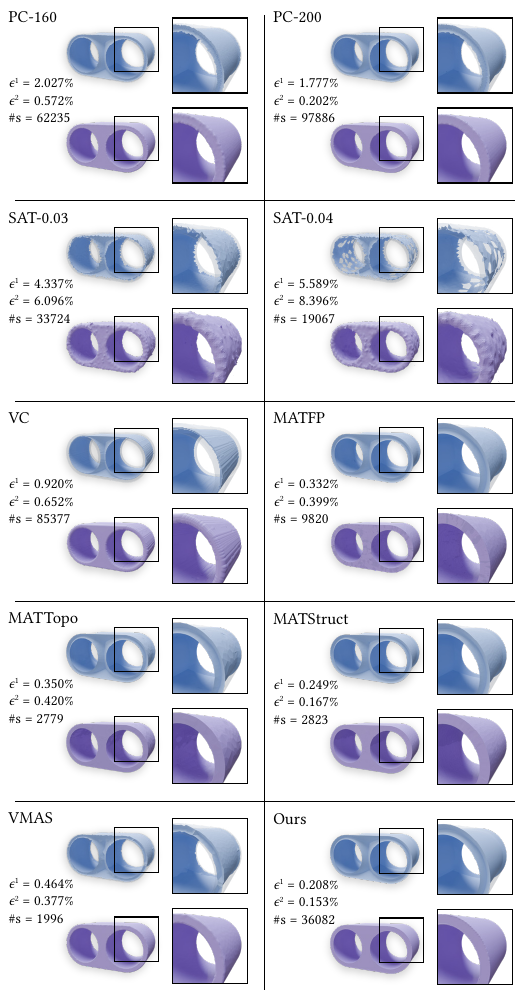}
\caption{Qualitative comparison of surface reconstruction on a genus-2 CAD model. The insets show zoomed-in details of the generated surfaces. We display the reconstruction error metrics ($\epsilon^1, \epsilon^2$) and the number of medial spheres ($\#s$). The reconstructed surfaces are rendered in \textbf{\textcolor{purple}{purple}}, while the underlying medial meshes are shown in \textbf{\textcolor{blue}{blue}}.}
\label{fig:geom_comparison}
\end{figure}

\subsection{Comparisons on CAD Shapes}
\label{subsec:cad}

Quantitative comparison with baseline methods over the 93 CAD models is shown in Tab.~\ref{tab:metrics_portrait}. We provide two sampling settings on meshes to provide point cloud input. To ensure consistent feature granularity across meshes of varying scales during point cloud conversion, we employ an adaptive sampling strategy based on surface area. For each input mesh $\mathcal{M}$, we first compute the maximum edge length $L_{\max}$ of its Axis-Aligned Bounding Box and its total surface area $S$. The target number of sample points $N$ is determined by a preset resolution ratio $R$. We define the target sampling stride as $\delta = L_{\max} / R$, which effectively divides the object's longest axis into $R$ units. Based on the relationship between the sampling stride and the surface area, the target point count is calculated as follows:

\begin{equation}
    N = \max \left( N_{\min}, \left\lfloor \frac{S}{\delta^2} \right\rfloor \right) = \max \left( N_{\min}, \left\lfloor S \cdot \left( \frac{R}{L_{\max}} \right)^2 \right\rfloor \right),
\end{equation}
where $N_{\min}$ serves as a minimum threshold to prevent under-sampling for extremely small or degenerate meshes. Subsequently, we utilize the Poisson Disk Sampling algorithm to generate $N$ points, ensuring a uniform distribution and minimizing aliasing artifacts on the surface. Specifically, for NSK, VMAS, and our method, we provide R = 160. For PC, we provide two options R = 160 and R = 200 to generate point samples. For oriented point cloud inputs, we generate uniformly distributed sample points from each mesh using a barycentric lattice scheme. By default, the total number of sample points is approximately 500,000, with slightly higher density in feature regions to ensure that sharp edges and corners are well captured. Each sample point is associated with a surface normal and a feature flag indicating whether it lies in a feature region, resulting in a complete oriented point cloud used for subsequent SQEM clustering and medial axis extraction. We use a voxel resolution of $2^8$ and a pruning parameter of lambda = 0.03 on VC, while our resolution of octree is $2^7$. Note that mesh inputs are used for SAT, VC, MATFP, MATTopo, and MATStruct. We set the SAT~\cite{miklos2010discrete} scaling factor as $s = 1 + \sigma$, where $\sigma$ represents the inflation ratio. In our experiments, we use $\sigma \in {0.03, 0.04}$. Since NSK is a neural-based skeleton method, we only evaluate its TQ and TER. 

To visually assess the topological robustness of our method, we compare the generated medial meshes on a CAD shape featuring sharp corners in Fig.~\ref{fig:topo_comparison}. As indicated by the Euler characteristic, methods that rely heavily on density estimation such as PC-160 and SAT frequently introduce topological noise, resulting in erroneous $\chi$ values. For instance, PC-160 generates a highly fragmented structure with $\chi=297694$, while SAT collapses the shape into disjoint components ($\chi=52$). In contrast, our method accurately preserves the input topology ($\chi=-6$), matching the topological correctness of VC, MATTopo, and MATStruct while maintaining a regular mesh structure.

Fig.~\ref{fig:geom_comparison} presents a comparison of geometric fidelity and surface quality on a genus-2 CAD shape. We report two one-sided Hausdorff distances $\epsilon^1$ and $\epsilon^2$ alongside the visual results. Volumetric approaches like SAT exhibit significant surface artifacts, resulting in high geometric errors ($\epsilon^1 > 4\%$). While MATStruct achieves a low error rate ($\epsilon^1 = 0.249\%$), our method outperforms all baselines with the lowest reconstruction error ($\epsilon^1 = 0.208\%$). As seen in the close-up views, our approach produces a smooth, close reconstructed surface that faithfully captures the curvature of the input, avoiding the discretization aliasing seen in VC or the surface gaps present in point-cloud-based methods.

Tab.~\ref{tab:metrics_portrait} presents a quantitative comparison between our method and state-of-the-art techniques across medial structural accuracy, mesh quality, topological correctness, and geometric fidelity. Overall, our method achieves a strong balance among these criteria, producing clean medial meshes with accurate structural classification, high element quality, near-zero topological error, and the best geometric fidelity in terms of Hausdorff distance.

In terms of medial structural accuracy, evaluated by the MSER metric, our method achieves an error rate of \textbf{3.0874\%}, ranking second among all compared methods and substantially outperforming most baselines. Although MATStruct obtains the lowest MSER of 0.9423\%, our method still maintains a low rate of misclassified seam and junction spheres. This demonstrates the effectiveness of our SQEM-based cell type estimation and downgrading strategy in preserving the correct medial structure, especially around singular transitions such as seams and junctions.

Regarding geometric fidelity, our method achieves the best HD value of \textbf{0.4203\%}, outperforming MATStruct (1.0918\%), MATTopo (1.2545\%), and SAT ($>4.4\%$). This indicates that our method better preserves local geometric details and sharp features, avoiding the geometric shrinkage or smoothing often observed in competing approaches.

Regarding element quality, evaluated via the TQ metric, our method achieves a score of \textbf{0.6831}, which is competitive with the best-performing baselines. While MATStruct and VMAS obtain slightly higher TQ scores of 0.7124 and 0.6953, respectively, VMAS exhibits much worse structural and geometric accuracy, with an MSER of 95.6633\% and an HD of 1.9718\%. This suggests that VMAS prioritizes element regularity at the expense of accurate medial structure and geometric fidelity. In contrast, our method maintains high element quality while preserving both the underlying geometry and the medial structure, providing a more balanced triangulation suitable for downstream applications.

In terms of TER, our method achieves a low error rate of \textbf{2.15\%}. This result is close to the zero TER reported by MATTopo and MATStruct, while VC obtains a TER of 3.0928\% on this dataset. Other baselines, including PC, SAT, NSK, and VMAS, show higher topological error rates under this evaluation. These results suggest that our method performs well in this empirical topological evaluation, although unlike theoretically guaranteed methods, our current framework does not provide a formal topological guarantee.

Regarding self-intersections, our method achieves an SITR of 4.0711\%. This value is substantially lower than SAT-0.03 (20.3429\%), SAT-0.04 (16.8332\%), and VMAS (10.4908\%), although it is higher than those of the Voronoi/RPD-based methods and Neural Skeleton. PC, MATFP, and MATTopo contain no self-intersecting triangles in this evaluation, while VC and MATStruct achieve near-zero SITRs of 0.0009\% and 0.0064\%, respectively. These low values are consistent with the fact that the connectivity of Voronoi/RPD-based methods is largely inherited from globally embedded Voronoi/Delaunay-type complexes. Neural Skeleton also achieves an SITR of 0: after selecting skeletal points through its coverage formulation, it constructs a regular triangulation of the selected skeletal points together with the surface samples and extracts the skeletal edges and triangles as a subcomplex of this triangulation. In contrast, SAT simplifies the medial structure through multiplicative scaling of medial balls and does not explicitly impose a non-self-intersection constraint on the resulting discrete medial mesh, which is consistent with its higher SITR in our evaluation.

For our method, the observed self-intersections mainly arise from the virtual-node mechanism used to recover local medial connectivity when an SQEM solution lies outside its current octree cell. If the reconstructed position lies in the current cell or a face-adjacent cell, the corresponding local Dual Contouring configuration preserves the usual face-based neighborhood relation, and we do not observe this type of self-intersection in our experiments. However, the virtual-node construction additionally allows a node to adopt an SQEM solution located in an edge-adjacent cell in order to recover difficult local configurations that would otherwise produce missing connections. This more permissive relocation weakens the strict geometric locality of the standard Dual Contouring construction: in some configurations, dual faces generated from neighboring cells can fold across one another, producing local self-intersections. Therefore, the current virtual-node strategy introduces a practical trade-off between recovering local medial connectivity and preserving a fully intersection-free embedding. Nevertheless, the measured SITR remains substantially lower than those of SAT and VMAS.

Overall, existing methods often optimize for a specific aspect, such as topology, element regularity, structural correctness, or geometric fidelity, while underperforming in others. Our method provides a balanced solution: it achieves the best Hausdorff distance, competitive MSER and triangle quality, and near-zero topological error. The SITR evaluation further shows that the method maintains a relatively low level of geometric self-intersection despite using the more permissive virtual-node construction to recover difficult local medial connections. These results demonstrate that our approach can generate clean and accurate medial meshes with robust performance across structural, geometric, and topological criteria.

\begin{table}[h]
\centering
\caption{Quantitative comparison on the organic dataset. Best values are shown in \textbf{bold}, second-best values are \underline{underlined}, and third-best values are shown in \textit{italics}.}
\label{tab:comparison}
\begin{tabular}{l|l|l|l|l|l}
\toprule
\footnotesize\textbf{Method}&\footnotesize\textbf{HD(\%)$\downarrow$}&\footnotesize\textbf{TER(\%)$\downarrow$} &\footnotesize\textbf{$TQ_{\text{avg}}$ $\uparrow$}&\footnotesize\textbf{$TQ_{\text{std}}$ $\downarrow$}&\footnotesize\textbf{SITR(\%)$\downarrow$}\\
\midrule
\textbf{Ours}& \textbf{0.775}&\underline{12.5}&\underline{0.653}& \textbf{0.0125}&13.6631\\
MATFP&3.532&\textit{75.0}&0.425&0.0242&\textbf{0}\\
MATTOPO&\underline{1.327}&\textbf{0.0}&0.515&\textit{0.0225} &\textbf{0}\\
MATStruct&3.183&\underline{12.5}&0.705&0.0735&\textbf{0}\\
NSK&--& 100.0&\textbf{0.678}&\underline{0.0184}&\textbf{0}\\
PC\_160&\textit{1.435}&100.0&0.395& 0.0253&\textbf{0}\\
PC\_200&1.770&100.0&0.384&0.0252&\textbf{0}\\
SAT-0.03&4.792&100.0&0.244&0.0256&13.8488\\
SAT-0.04&6.186&100.0&0.235&0.0253&11.3623\\
VC&2.274&\textbf{0.0}&0.424&0.0241&\underline{0.0006}\\
VMAS&1.623&100.0&\textit{0.578}&0.0238&\textit{8.8253}\\
\bottomrule
\end{tabular}
\end{table}

\begin{figure}
\centering
\includegraphics[width=0.94\linewidth]{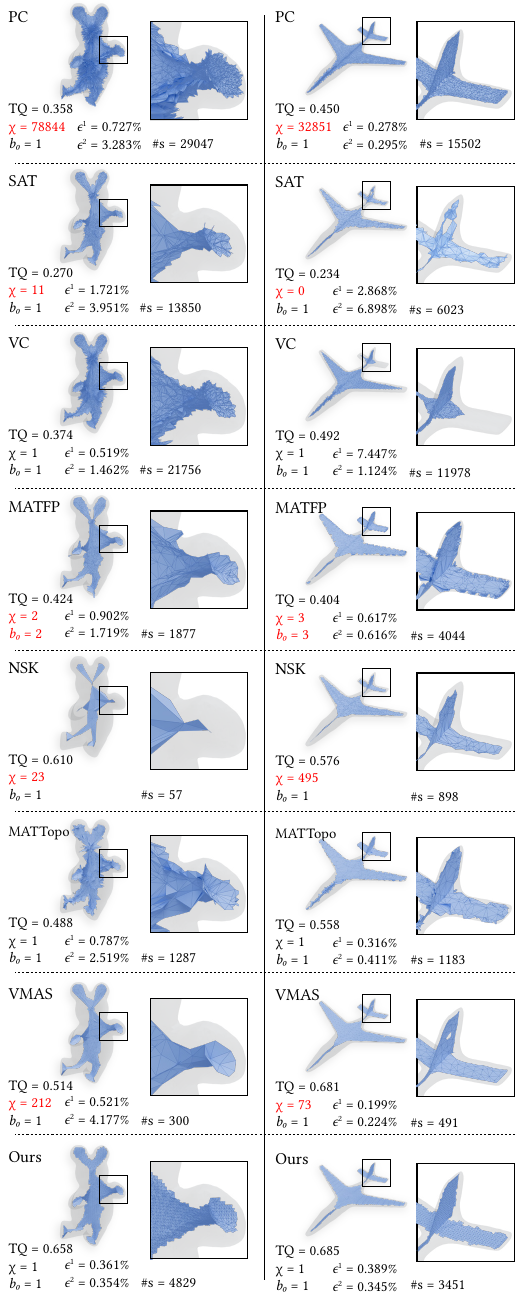}
\caption{
Visual comparison on organic shapes (Bear and Plane). We report the Triangle Quality (TQ), Euler characteristic ($\chi$), reconstruction errors ($\epsilon^1$ and $\epsilon^2$), and number of medial balls ($\#s$) for each reconstruction. Topological errors are highlighted in \textbf{\textcolor{red}{red}}.
}
\label{fig:ncad_comparison}
\end{figure}

\subsection{Direct Extraction vs. Surface-Reconstruction Pipelines}
\label{subsec:reconstruction_pipeline}

To further evaluate the practical benefit of directly extracting a medial mesh from an oriented point cloud, we compare our method with two reconstruction-based pipelines that first reconstruct a surface, tetrahedralize the reconstructed surface, and then apply MATStruct. Screened Poisson is evaluated using exactly the same oriented point-cloud input as our method. For RFEPS, we additionally consider a second point-cloud construction and retain the better reconstruction for each model, as detailed below. This experiment differs from the comparison in Tab.~\ref{tab:metrics_portrait}, where MATStruct is directly provided with the original high-quality mesh. Here, the goal is to evaluate the complete processing pipeline required before medial-axis extraction when the input is represented as an oriented point cloud.

The first pipeline uses RFEPS~\cite{xu2022rfeps}, a feature-preserving surface reconstruction method designed for CAD-like inputs. We evaluate RFEPS using two point-cloud constructions. In the first setting, RFEPS is given exactly the same sampled oriented point cloud used as input to our method. In the second setting, we directly use the vertices of the original triangular mesh together with their normals, thereby avoiding possible effects introduced by our point-sampling procedure. For each model, we inspect the reconstructions obtained under both settings and retain the better result based on visual reconstruction quality and topological consistency with the ground-truth surface. The aggregate RFEPS results reported below therefore represent a per-model selection between the two settings. To avoid conflating feature-detection errors with surface-reconstruction performance, we also bypass the feature-detection stage and directly provide the available feature-point indices. This per-model selection and the directly supplied feature labels deliberately provide a favorable evaluation setting for RFEPS.

Even after this favorable per-model selection, only 47 of the 93 selected RFEPS reconstructions have the same topology as the corresponding ground-truth surfaces. We emphasize that a topology mismatch at the reconstructed-surface stage does not necessarily propagate to the subsequent stages. For example, some reconstructed surfaces contain only small local holes, while after tetrahedralization the resulting volumetric mesh can still recover the same global topology as the ground truth. We therefore do not directly equate a surface-level topology error with failure of the final medial-mesh pipeline.

Nevertheless, we also observe clearly incomplete RFEPS reconstructions. Some reconstructed surfaces contain numerous large holes, while in more severe cases only a small portion of the original surface is recovered, which we regard as a surface-reconstruction failure. We do not further speculate on the specific algorithmic causes of these failures. In addition, some reconstructed surfaces cannot be successfully tetrahedralized, and a smaller number of failures can still occur during the subsequent MATStruct extraction. After the complete RFEPS--tetrahedralization--MATStruct pipeline, medial meshes are successfully produced for 67 of the 93 models, of which 63 have the correct topology.

The second pipeline employs Screened Poisson surface reconstruction~\cite{kazhdan2013screened}, which is substantially more robust in terms of reconstruction success. All 93 CAD models are successfully reconstructed, and 89 of the reconstructed surfaces preserve the topology of the original shape. However, Screened Poisson does not explicitly preserve the sharp CAD features that are important for the feature-preserving medial structures considered in our work. We subsequently tetrahedralize each reconstructed surface and apply MATStruct using the same procedure as in our mesh-based baseline evaluation. Additional failures can occur during tetrahedralization or medial-mesh extraction: the complete Screened-Poisson--MATStruct pipeline successfully produces medial meshes for 86 models, of which 83 have the correct topology.

\begin{table}[htbp]
    \centering
    \caption{
    Comparison between direct medial-mesh extraction and reconstruction-based pipelines on 93 CAD models extracted from ABC dataset. For each reconstruction-based pipeline, the number of successfully generated final medial meshes is reported separately, while MSER, TQ, TER, HD, and SITR are evaluated only over these successful outputs. Pipeline failures during surface reconstruction, tetrahedralization, or medial-mesh extraction are therefore not counted as topology errors. $N_{\mathrm{eval}}$ denotes the number of models for which the complete pipeline successfully produces a final medial mesh and that are included in the metric evaluation. For RFEPS, the reported results are obtained through the per-model selection between the two input constructions described above.
    }
    \label{tab:reconstruction_pipeline}
    \setlength{\tabcolsep}{5pt}
    \renewcommand{\arraystretch}{1.25}

    \begin{tabular}{l|l|l|l|l|l|l}
        \hline
        \footnotesize\textbf{Pipeline}
        &
        {\footnotesize\textbf{$N_{\mathrm{eval}}$}}
        & \footnotesize\textbf{MSER(\%)$\downarrow$}
        & \footnotesize\textbf{TQ$\uparrow$}
        & \footnotesize\textbf{TER(\%)$\downarrow$}
        & \footnotesize\textbf{HD(\%)$\downarrow$}
        & \footnotesize\textbf{SITR(\%)$\downarrow$} \\
        \hline

        \textbf{Ours}
        & 93
        & 3.0874
        & 0.683
        & 2.1505
        & 0.4203
        & 4.0711 \\
        \hline

        Poisson
        & 86
        & 3.5349
        & 0.733
        & 3.488
        & 1.4854
        & 0 \\

        RFEPS
        & 67
        & 1.9255
        & 0.724
        & 5.970
        & 1.3922
        & 0 \\
        \hline
    \end{tabular}
\end{table}

Table~\ref{tab:reconstruction_pipeline} summarizes the quantitative results of the complete pipelines. Our method produces a final medial mesh for all 93 CAD models, whereas the Poisson+MATStruct and RFEPS+MATStruct pipelines produce final outputs for 86 and 67 models, respectively. Even with the favorable per-model selection used for RFEPS, this difference illustrates that the robustness of a reconstruction-based pipeline depends not only on the medial-axis extraction method itself, but also on the preceding surface-reconstruction and tetrahedralization stages.

Among the successfully generated outputs, our method achieves the lowest TER of 2.1505\%, compared with 3.4884\% for Poisson+MATStruct and 5.9701\% for RFEPS+MATStruct. It also achieves the lowest average HD by a substantial margin, 0.4203\%, compared with 1.4854\% and 1.3922\%, respectively. RFEPS+MATStruct obtains the lowest MSER of 1.9255\%, while Poisson+MATStruct achieves the highest average triangle quality of 0.733. These results indicate that no single approach dominates every metric. Nevertheless, direct extraction provides the highest pipeline success rate and, on average, the lowest topological error and geometric deviation in this experiment.

\begin{figure}[t]
    \centering
    \includegraphics[width=\linewidth]{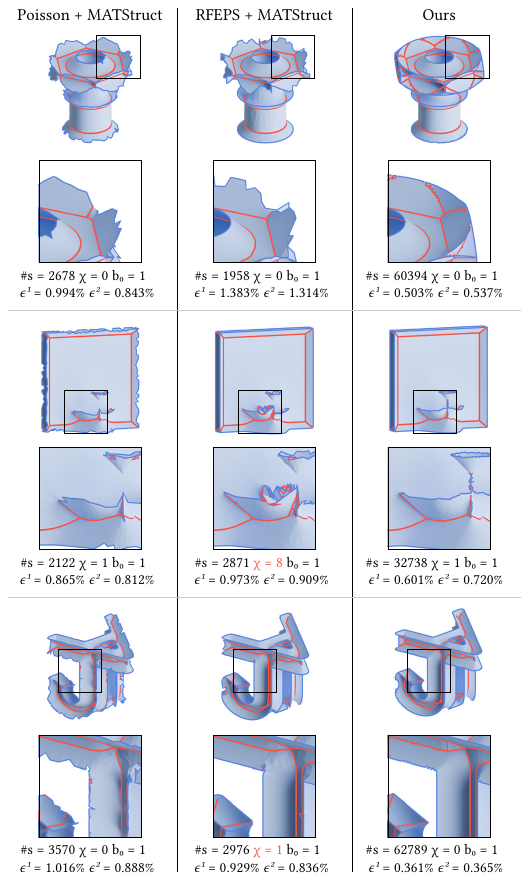}
    \caption{
    Visual comparison of Poisson+MATStruct, RFEPS+MATStruct, and our method on three representative CAD models. Each result is accompanied by a local zoom-in view. We report the number of medial spheres ($\#s$), Euler characteristic ($\chi$), number of connected components ($b_0$), and the two directional Hausdorff errors ($\epsilon^1$ and $\epsilon^2$).
    }
    \label{fig:reconstruction_pipeline_visual}
    \vspace{1.5pt}
\end{figure}

\enlargethispage{1.15\baselineskip}

The comparison also reveals a trade-off with respect to self-intersections. Both reconstruction-based pipelines have zero SITR on their successful outputs, whereas our method has an SITR of 4.0711\%. As discussed in Sec.~\ref{subsec:cad}, these self-intersections mainly arise from the virtual-node relaxation used by our Dual Contouring construction, particularly when a virtual node refers to a solution in an edge-adjacent cell. We retain this relaxation because restricting virtual nodes more aggressively can instead lead to missing local medial connections. Thus, the direct approach improves pipeline availability, topology recovery, and average geometric fidelity in this comparison, but does not dominate the reconstruction-based pipelines in every aspect.

Fig.~\ref{fig:reconstruction_pipeline_visual} provides a more detailed qualitative comparison on three representative models. For the first model, all three approaches produce a single connected medial complex with $\chi=0$ and $b_0=1$. Poisson+MATStruct obtains $\epsilon^1=0.994\%$ and $\epsilon^2=0.843\%$, RFEPS+MATStruct obtains $1.383\%$ and $1.314\%$, and our method obtains $0.503\%$ and $0.537\%$. Thus, while all three results share the same reported global topological indicators, our method achieves the smallest directional Hausdorff errors on this example. The local close-ups also reveal visible differences in the recovered medial-sheet geometry near the highlighted feature region.

The second model more clearly illustrates the interaction between geometric and topological quality. Poisson+MATStruct and our method both produce a single connected result with $\chi=1$ and $b_0=1$, whereas RFEPS+MATStruct produces $\chi=8$ with $b_0=1$. The local zoom-in of the RFEPS-based result contains substantially more complicated medial structures around the highlighted region, consistent with the large change in Euler characteristic. Nevertheless, its directional Hausdorff errors remain relatively small, with $\epsilon^1=0.973\%$ and $\epsilon^2=0.909\%$, illustrating that geometric proximity alone does not necessarily imply agreement in global topological indicators. Poisson+MATStruct obtains $\epsilon^1=0.865\%$ and $\epsilon^2=0.812\%$, while our method achieves the smallest errors in this example, with $\epsilon^1=0.601\%$ and $\epsilon^2=0.720\%$, while matching the reported global topological indicators of Poisson+MATStruct.

A similar distinction appears in the third model. Poisson+MATStruct and our method both have $\chi=0$ and $b_0=1$, whereas RFEPS+MATStruct has $\chi=1$ and $b_0=1$. RFEPS+MATStruct obtains $\epsilon^1=0.929\%$ and $\epsilon^2=0.836\%$, and Poisson+MATStruct obtains $1.016\%$ and $0.888\%$. Our method achieves the smallest directional Hausdorff errors, with $\epsilon^1=0.361\%$ and $\epsilon^2=0.365\%$. The close-up further shows that the three pipelines recover visibly different local medial-sheet structures near the highlighted feature region. In this example, our result therefore provides both lower geometric error and the same reported global topological indicators as Poisson+MATStruct.

Taken together, these visual examples are consistent with the aggregate results in Tab.~\ref{tab:reconstruction_pipeline}. Our method attains the smallest directional Hausdorff errors in all three examples while matching the reported global topological indicators of Poisson+MATStruct. In contrast, RFEPS+MATStruct exhibits substantial changes in Euler characteristic in the second and third examples despite maintaining relatively small geometric errors, indicating that surface-reconstruction errors can propagate to the subsequent medial extraction stage without necessarily producing large Hausdorff discrepancies. Screened Poisson produces more consistent global topological indicators in these examples, but it does not explicitly preserve sharp CAD features. Our method instead operates directly on the oriented point cloud and avoids both surface reconstruction and tetrahedralization, thereby removing two intermediate stages at which geometry, topology, or pipeline robustness may be affected.

We emphasize that this experiment is not intended to suggest that MATStruct itself is insufficiently robust. When MATStruct is directly provided with the original high-quality volumetric mesh, the results in Tab.~\ref{tab:metrics_portrait} demonstrate its strong topological and geometric quality. Rather, this experiment evaluates the additional preprocessing and potential failure modes introduced when only oriented point-cloud input is available. Under the clean and sufficiently dense CAD sampling considered here, direct extraction provides a practical advantage in overall pipeline success rate and average topological and geometric fidelity. Reconstruction-based pipelines nevertheless remain attractive when a compact medial representation is preferred and, as shown by our aggregate evaluation, may retain advantages in other representation-dependent properties such as self-intersection behavior.

\subsection{Comparisons on Organic Shapes}
\label{subsec:ncad}
We evaluate our method on 8 organic shapes and compare it with the same set of baselines used in the CAD experiments. We use the same input sampling, pre-processing, and experimental settings as those of CAD models. Table.~\ref{tab:comparison} presents quantitative metrics, including Hausdorff distance (HD), Topological Error Rate (TER), average triangle quality ($TQ_{\text{avg}}$), Self-Intersecting Triangle Ratio (SITR), and triangle quality standard deviation ($TQ_{\text{std}}$).

From Table.~\ref{tab:comparison}, our method achieves the lowest HD (\textbf{0.775\%}) and the lowest $TQ_{\text{std}}$ (\textbf{0.0125}) on the organic dataset, suggesting that it produces geometrically accurate results with relatively stable triangle quality. It also obtains a TER of 12.5\%, which is the same as MATStruct and remains relatively low on this small set of examples, although MATTopo and VC achieve zero topological error. The average triangle quality of our method ($TQ_{\text{avg}}=0.653$) is competitive, although MATStruct and NSK achieve higher values of 0.705 and 0.678, respectively. MATStruct achieves the highest average triangle quality on this dataset, but its $TQ_{\text{std}}$ of 0.0735 is considerably higher than ours, indicating a less uniform distribution of triangle quality. Its HD of 3.183\% is also substantially higher than the 0.775\% achieved by our method.

Regarding self-intersections, our method obtains an SITR of 13.6631\% on the organic dataset, which is higher than its SITR on the CAD dataset. Organic shapes generally exhibit more irregular local medial configurations across the octree cells. Consequently, SQEM solutions fall outside their current cells more frequently, leading to more frequent use of virtual nodes. The increased number of virtual nodes raises the likelihood of local face folding and self-intersections. SAT-0.03, SAT-0.04, and VMAS also exhibit relatively high SITRs of 13.8488\%, 11.3623\%, and 8.8253\%, respectively, whereas the remaining methods have zero or near-zero SITRs in this evaluation.

Overall, these results indicate that our method performs favorably on the tested organic models, particularly in terms of geometric fidelity and triangle-quality consistency, while self-intersections remain a limitation on free-form shapes with more irregular local medial configurations. We note that this dataset contains only eight shapes and therefore mainly serves as a qualitative indication rather than a comprehensive statistical benchmark.

Fig.~\ref{fig:ncad_comparison} compares different medial axis extraction methods on two representative non-CAD models. Each column corresponds to one model, and each row shows the result of one method together with a local zoom-in view. Existing methods exhibit different limitations on these organic shapes. PC produces very dense and noisy medial structures with extremely large Euler characteristic values. SAT and VC generate smoother results, but the extracted sheets may become fragmented or locally inaccurate. MATFP and MATTopo produce cleaner surfaces, but some local structures are still simplified or disconnected. NSK and VMAS generate compact skeleton-like structures, but they tend to lose sheet-level details and show larger topological discrepancies on these examples. In contrast, our method produces clean and coherent medial sheets for both models. It achieves high triangle quality, with $TQ=0.658$ and $TQ=0.685$, while maintaining $\chi=1$ and $b_0=1$ in this empirical evaluation.

\subsection{Implementation Details and Parameter Sensitivity}
\label{subsec:study}
\begin{figure}
\centering
\includegraphics[width=\linewidth]{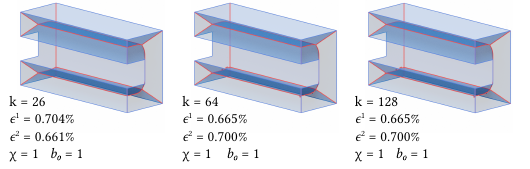}
\caption{\label{fig:knn_sensitivity}Impact of KNN parameter $K$ on projection grouping and medial sheet connectivity. Three values of $K$ (26, 64, 128) are tested on a CAD model. The resulting medial sheets remain topologically correct and consistent across all $K$ values, indicating robustness of the method to the choice of $K$. Normal-based edge filtering removes edges across sharp features, ensuring stable clustering even with larger $K$ values. The corresponding Hausdorff errors ($\epsilon^1$, $\epsilon^2$) and Euler characteristic ($\chi$) are also reported for each $K$.}
\end{figure}

\paragraph{KNN Graph Construction}
To cluster projected points and determine connectivity in the medial axis extraction, we build a $K$-nearest neighbor (KNN) graph. We evaluate $K \in \{26,64,128\}$ to analyze parameter sensitivity. As shown in Fig.~\ref{fig:knn_sensitivity}, even with a large value such as $K=128$, the resulting medial sheets remain correct, indicating that our method is not sensitive to this parameter within a reasonable range. The main reason is that the KNN graph is not used directly: during graph construction, we remove edges between two non-feature points when their normal angle exceeds $45^\circ$, which corresponds to the complement of our dihedral-angle threshold for feature-edge detection. For CAD models, the most common problematic KNN edges are those that cross a sharp feature and connect two surface regions that should be disconnected. Our normal-based filtering effectively suppresses such cross-feature edges, making the subsequent projection grouping stable even when a larger $K$ introduces many additional candidate edges.

\begin{figure}
\centering
\includegraphics[width=\linewidth]{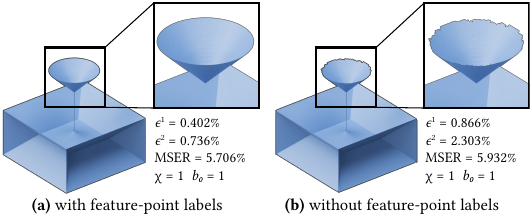}
\caption{
\label{fig:ablation_feature}
Effect of removing feature‐point labels. (a) With feature‐point labels, the medial surface tightly follows the sharp boundary. (b) Without these labels, the medial surface near the outer rim becomes visibly eroded and incomplete, leading to a larger Hausdorff error; the MSER shows only a slight degradation, while the Euler characteristic remains identical across both results.
}
\end{figure}

\begin{figure}[t]
    \centering
    \includegraphics[width=\linewidth]{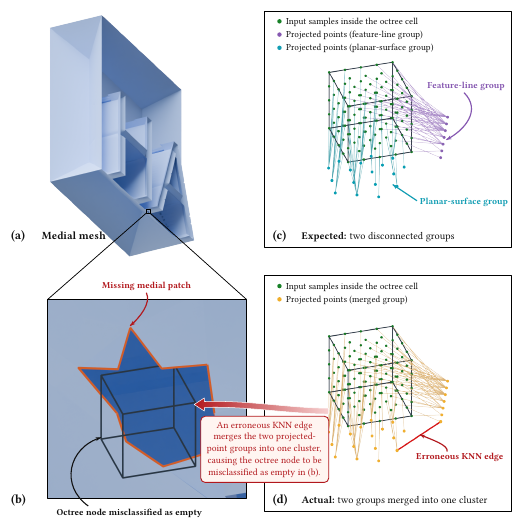}
    
    \caption{
    A failure case of KNN-based connectivity on a CAD model.
    (a) The extracted medial mesh is shown with non-manifold edges highlighted in red; the black box marks the problematic region.
    (b) A close-up shows the missing medial patch and the octree node incorrectly classified as empty.
    (c) Expected connectivity: the projected points from the feature line and the adjacent planar region form two separate groups.
    (d) Actual connectivity: a spurious KNN edge links the two projected-point groups, merging them into one cluster. The octree node is consequently classified as empty, so the corresponding medial patch is omitted and a local topological error results.
    }
    \label{fig:knn_failure}
\end{figure}

\paragraph{Failure Case of KNN-based Connectivity.}
Although the normal-based filtering described above suppresses most spurious KNN connections in our experiments, the resulting graph still provides only a discrete approximation of the underlying surface connectivity and therefore does not provide a theoretical guarantee. Fig.~\ref{fig:knn_failure} shows a representative failure case from our CAD benchmark. The extracted medial mesh contains a local hole, resulting in incorrect topology.

The zoomed-in view reveals the source of this failure. For the highlighted octree node, the projected surface samples should form two distinct groups: one associated with a feature line and the other with a planar surface region. However, an erroneous KNN edge remains between a pair of samples belonging to these two regions after graph filtering. Consequently, the two projection groups become connected in the KNN graph and are merged into a single group. The node is therefore incorrectly classified as empty, causing the corresponding medial structure to be missed and ultimately producing the hole in the extracted medial mesh. This example illustrates the heuristic nature of the KNN-based connectivity approximation: while it is effective for the sampling regimes considered in most of our experiments, an incorrect graph connection can still merge geometrically distinct surface regions and lead to a topological error.
% \begin{figure}
% \centering
% \includegraphics[width=\linewidth]{pics/ablation-rank.pdf}
% \caption{\label{fig:ablation_rank}
% \xiaohu{There is no such "feature-preserving strategy" mentioned in the paper now! This figure needs to be either modified or removed!} Effect of the feature-preserving strategy based on ($s_{\text{fea}}$) and the geometric priority strategy based on ($s_{\text{geo}}$).
% (a) Result with both ($s_{\text{geo}}$) and ($s_{\text{fea}}$) enabled (ours), where the medial sheet closely follows the internal feature line (highlighted in \textbf{red}) and aligns well with the outer boundary (shown in \textbf{dark blue}), achieving the lowest Hausdorff errors ($(\varepsilon_1,\varepsilon_2)$) and MSER.
% (b) Without the geometric priority ($s_{\text{geo}}$), the medial sheet near the red internal feature line becomes wavy and misaligned in the zoom-in view, and all error metrics increase.
% (c) Without the feature-preserving term ($s_{\text{fea}}$), a visible tear appears along the dark-blue outer boundary (zoom-in), leading to a further increase in Hausdorff error and MSER.
% In all three settings, the global topology (all three Euler characteristic measurements) remains identical.
% }
% \end{figure}

\begin{figure}
\centering
\includegraphics[width=\linewidth]{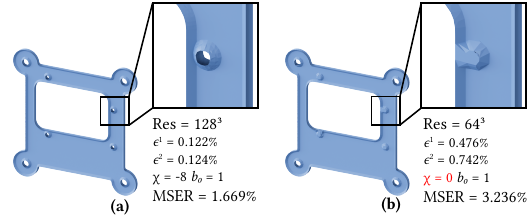}
\caption{
\label{fig:ablation_resolution}
Comparison of medial results under different octree resolutions. (a) With resolution ($128^3$), the medial surface is geometrically and topologically accurate. (b) At a coarser resolution of ($64^3$), both the Hausdorff error and MSER increase noticeably, and in the zoomed region a hole that should remain open is incorrectly sealed by a spurious medial patch.
}
\end{figure}

\begin{figure}
\centering
\includegraphics[width=\linewidth]{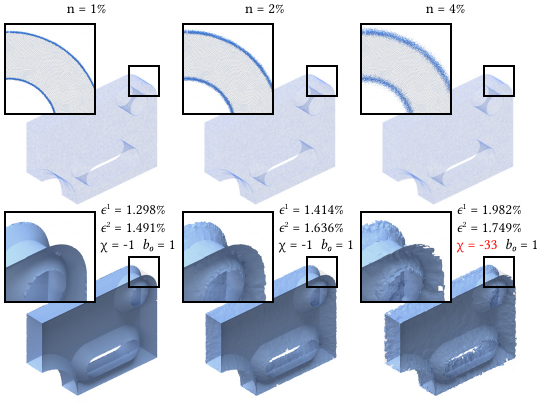}
\caption{\label{fig:noise-comparison}\textbf{Robustness to noise.} From left to right, input point clouds are corrupted with Gaussian noise of magnitude 1\%, 2\%, and 4\% of the minimum side length of the bounding box. The top row shows the noisy oriented point clouds with local zoom-ins of the point cloud and underlying shape; the bottom row shows the corresponding medial meshes, where higher noise levels lead to increasingly noisy medial sheets with incorrect topology.}
\end{figure}

\begin{figure}
\centering
\includegraphics[width=\linewidth]{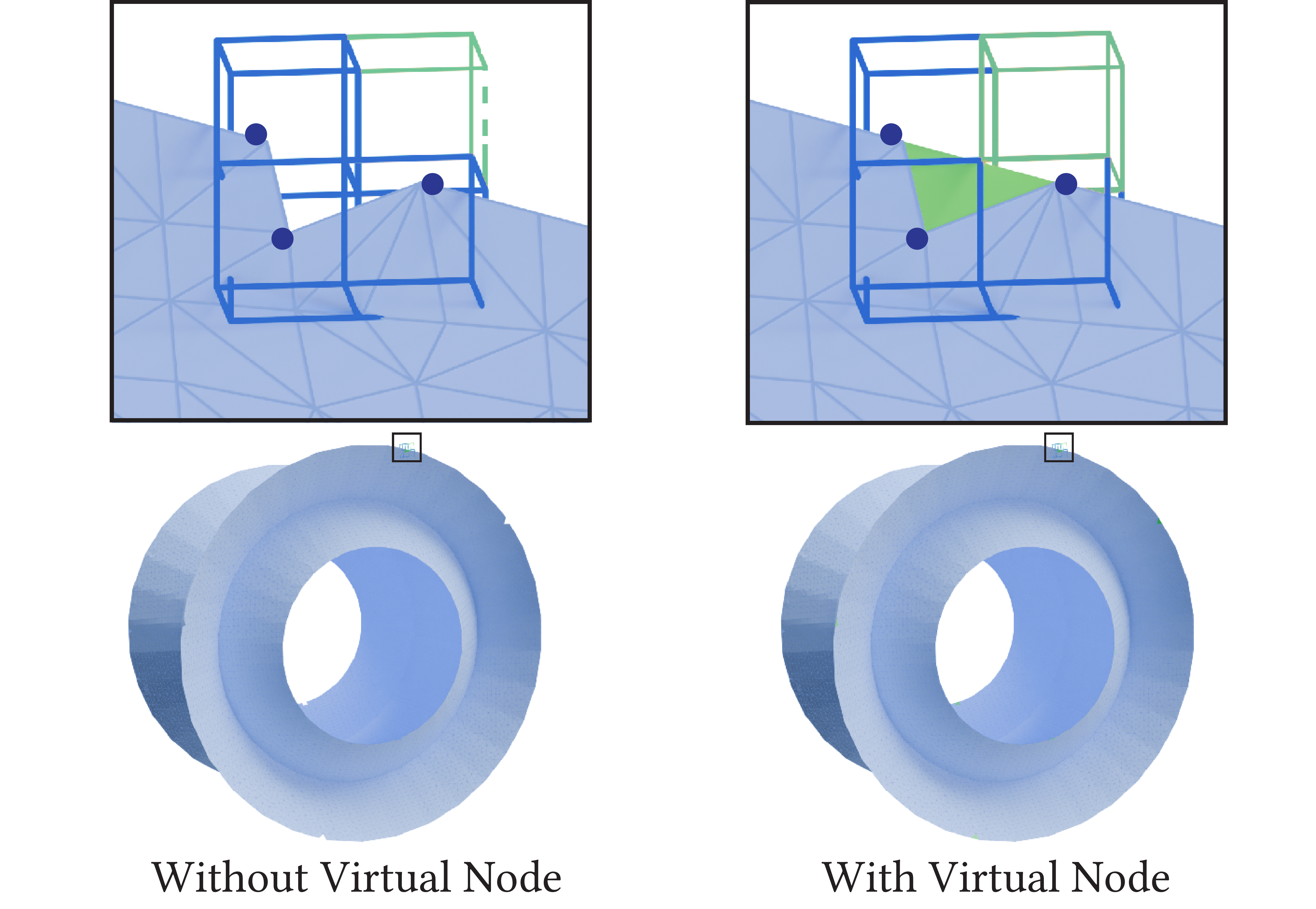}
\caption{\label{fig:virtual-outer-feature}Using a virtual node (green cell) near the outer boundary to complete the four-cell stencil of a grid edge. Without the virtual node, the corresponding green face on the medial mesh would be missing, creating an open gap. }
\end{figure}

\begin{figure}
\centering
\includegraphics[width=\linewidth]{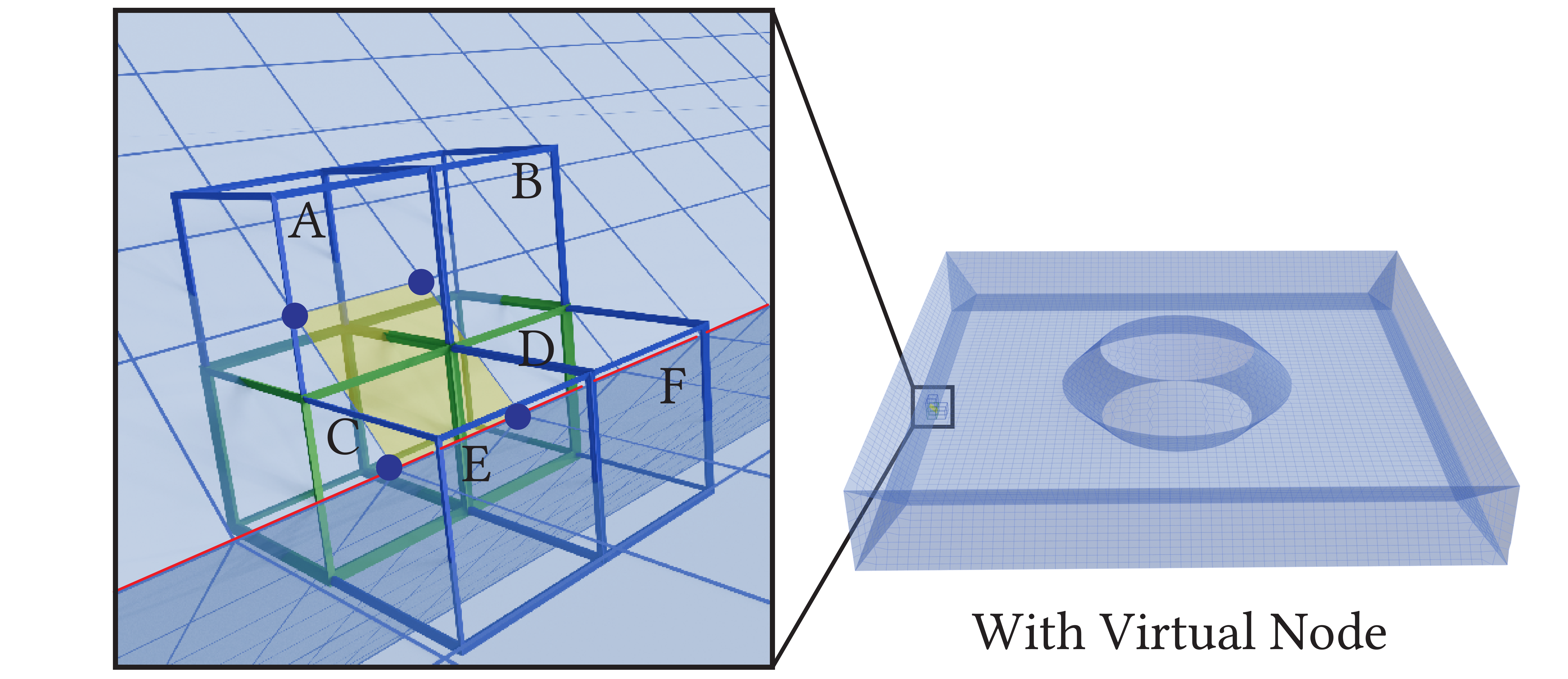}
\caption{\label{fig:virtual-inner-feature}Virtual nodes near a medial seam (red line) region. Dark blue cubes denote cells of real nodes (A, B, E, F), green cubes are cells of virtual nodes (C, D), and the yellow quad constructed by dual contouring is geometrically degenerate because two of its virtual nodes copy the positions of neighboring real nodes on a medial seam (C←E, D←F). We keep such quads to maintain correct topological connectivity and remove them later during thinning and triangulation.}
\end{figure}

\paragraph{Effect without Feature-point Labels.}
When feature point labels are provided as input, for any non-empty octree cell containing such points, we directly select the geodesic centroid of these feature points as its dual vertex instead of relying on the SQEM solution. In the ``no feature labels'' configuration, if the SQEM solution of a non-empty cell is classified as a Corner or Line type (Sec.~\ref{subsec:sqem}), we treat its dual vertex as a feature vertex. A visual comparison of the two setups is shown in Fig.~\ref{fig:ablation_feature}.
We observe that, without feature labels, medial sheets near the outer boundary exhibit noticeable missing regions. The reason is that in each cell we only use 27 samples to construct the SQEM system and discard samples whose SDF value is greater than zero. Near the outer boundary, the number of valid samples can easily drop to zero, making SQEM fitting impossible. Such cells are then marked as empty, which effectively "hollows out" the medial surface along the boundary.

\begin{figure}[t]
    \centering
    \includegraphics[width=\linewidth]{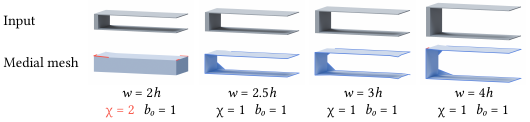}
    \caption{
    Effect of the separation between nearby surface sheets relative to the local voxel size $h$. Each thin layer has a thickness of $0.125h$, while the gap width is varied as $w=2h$, $2.5h$, $3h$, and $4h$. The $2h$ case produces incorrect topology, whereas the other three cases recover the correct topology. The Euler characteristic $\chi$ and the number of connected components $b_{0}$ are reported for each result. Non-manifold edges in the medial meshes are highlighted as \textbf{\textcolor{red}{red}} lines.
    }
    \label{fig:ablation_lfs}
\end{figure}

\paragraph{Octree Resolution and Local Feature Size.}
We ablate the influence of the maximum octree depth, as shown in Fig.~\ref{fig:ablation_resolution}. When the resolution is as low as $64^3$, some regions that should contain small holes or thin tunnels are mistakenly bridged and closed by a single quadrilateral face. Our topology-aware thinning only removes faces that belong to detected cubic solid-cell complexes, and this spurious face does not participate in any of such solid-cells; as a result, it survives the simplification, revealing an inherent limitation of our DC-SQEM when the octree resolution is too low.

We further examine the relation between the local voxel size $h$ and the local feature size (LFS) using a controlled narrow-gap example, as shown in Fig.~\ref{fig:ablation_lfs}. Our experiments indicate that the minimum LFS alone does not determine the required voxel resolution. In the synthetic example, each thin layer has a thickness of only $0.125h$, corresponding to an interior LFS of approximately $0.0625h$ away from its boundary edges, yet the medial sheets of the thin layers can still be successfully recovered. This observation is also consistent with our CAD benchmark: among the 93 CAD models, five contain very thin structures, whose medial sheets are also successfully reconstructed.

We therefore focus on the LFS associated with the separation between distinct nearby surface sheets. Let $w$ denote the gap width between the two layers. For approximately parallel surfaces, the corresponding surface local feature size is approximately $\mathrm{LFS}_{\mathrm{gap}}\approx w/2$. We test four gap widths, $w=2h$, $2.5h$, $3h$, and $4h$. As shown in Fig.~\ref{fig:ablation_lfs}, the case $w=2h$ ($\mathrm{LFS}_{\mathrm{gap}}\approx h$) produces incorrect topology, whereas the cases $w=2.5h$, $3h$, and $4h$ ($\mathrm{LFS}_{\mathrm{gap}}\approx1.25h$, $1.5h$, and $2h$, respectively) all produce the correct topology. The Euler characteristic and the number of connected components are reported for each case in the figure.

This difference is reflected by the corresponding octree-node configurations. For $w=2h$, the non-empty nodes associated with the two nearby layers remain continuously adjacent across the gap, resulting in an erroneous connection. For the larger gaps tested here, empty nodes appear between the two groups of non-empty nodes and separate the corresponding medial structures. These observations indicate that the minimum LFS of the shape is not by itself a resolution criterion for DC-SQEM; rather, for this type of failure, the more relevant scale is the LFS associated with the separation between distinct nearby surface sheets.

\paragraph{Robustness to Noise.} We evaluate the impact of noise on medial axis computation. Starting from the same CAD model, we generate three oriented point clouds with labeled feature points and add Gaussian noise with standard deviations equal to 1\%, 2\% and 4\% of the minimum side length of the bounding box, respectively. As shown in Fig.~\ref{fig:noise-comparison}, higher noise levels lead to increasingly irregular medial boundaries near the outer surface, and the measured Hausdorff Distance increases accordingly; the topology of the computed medial mesh also exhibits deterioration as the noise level is increased to 4\%.

\begin{figure}[htbp]
    \centering
    \includegraphics[width=\linewidth]{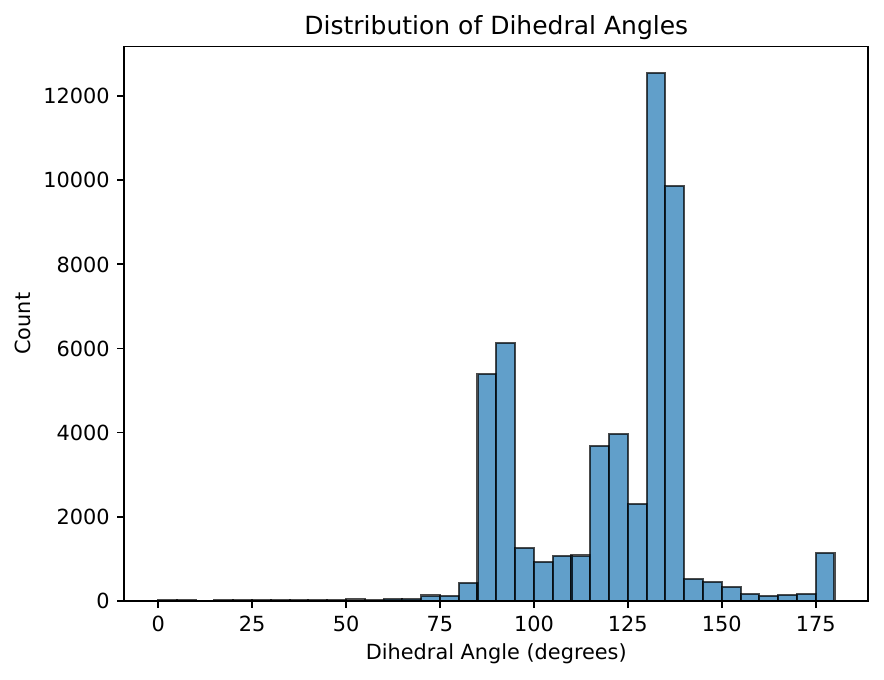}
    \caption{
    Histogram of dihedral angles for medial faces associated with internal feature lines, collected from 88 CAD models and 52{,}173 samples. 
    %Only a very small fraction ($\sim$0.56\%) of angles fall below the threshold of $70.53^\circ$, showing that the failure case is extremely rare in practice.
    }
    \label{fig:virtual-node-angle-hist}
\end{figure}

\subsection{Discussion of Virtual Nodes}
\label{sec:discussion_virtualnodes}
We analyze the role of virtual nodes in preserving local connectivity during dual contouring. Figs.~\ref{fig:virtual-outer-feature} and~\ref{fig:virtual-inner-feature} each show a full model together with a zoom-in view of a representative local configuration. Fig.~\ref{fig:virtual-outer-feature} illustrates a typical boundary case: dark blue cubes denote real nodes, while the green cube is a virtual node that references a medial sphere in an adjacent real node. If only real nodes are used, the four cells around the corresponding grid edge are incomplete, so no quad face can be generated and a gap appears near the outer boundary. By treating the green cell as non-empty and inserting a virtual node, the four-cell stencil becomes complete, allowing dual contouring to generate a full quad on the medial mesh (with two vertices sharing the same position) and fill the hole.

Fig.~\ref{fig:virtual-inner-feature} shows a more complex example near an internal feature region. In this case, six nodes meet locally: four blue nodes ($A$, $B$, $E$, and $F$) are real, and two green nodes ($C$ and $D$) are virtual. Two quad faces around this region share one edge. The quad highlighted by the yellow face is geometrically valid and non-degenerate, since its four vertices come from four neighboring cells ($A$ and $B$ real, $C$ and $D$ virtual). The other quad, however, is geometrically degenerate: because the virtual nodes $C$ and $D$ copy the coordinates of their referenced real nodes $E$ and $F$, this second quad collapses to a single edge of the yellow quadrilateral. Without the two virtual nodes $C$ and $D$, the yellow quad could not be formed and the medial mesh would contain a hole at this location. Therefore, virtual nodes are introduced to preserve the correct topological connectivity first, while the resulting degenerate quads are safely resolved in later stages through vertex merging and face cleanup.

\begin{figure}
\centering
\includegraphics[width=\linewidth]{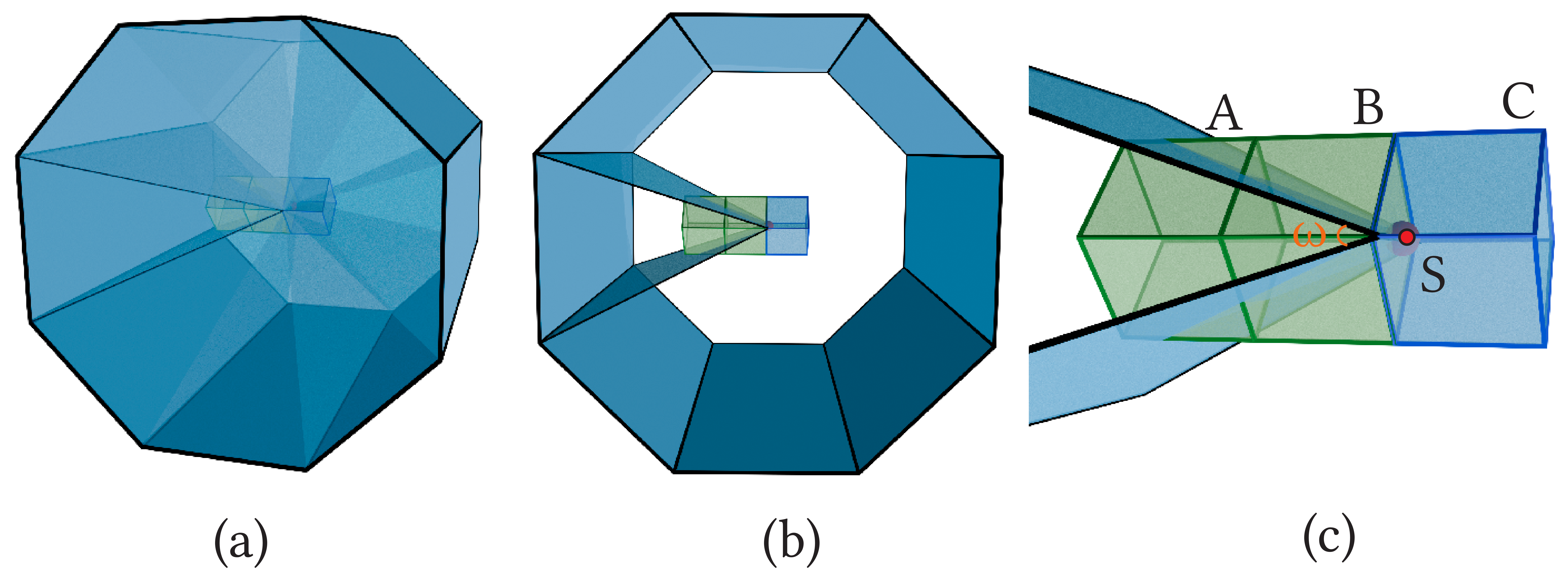}
\caption{\label{fig:virtual-node-limitation}A constructed example illustrating the limitation of virtual nodes. (a) A regular octagonal prism with caps and its medial axis. (b) A cross-sectional view of the same configuration. (c) A zoom-in of the local region in (b), where only two medial sheets are shown for clarity. The two sheets pass through two vertices of cell \(A\) and intersect another medial sheet along a line passing through cell \(C\), which is two grid units away from \(A\). In this configuration, the required virtual reference lies beyond the face- or edge-adjacent neighborhood, so the current virtual-node mechanism cannot recover the missing local connection.}
\end{figure}

Currently, we only introduce virtual nodes between face-adjacent or edge-adjacent octree leaf nodes to preserve local topological connectivity, so the effect of virtual nodes is inherently local. We manually constructed a counterexample (see Fig.~\ref{fig:virtual-node-limitation}) to illustrate the failure of virtual-node mechanism: two medial sheets pass exactly through two vertices of a cell $A$, and both sheets intersect another medial sheet along a line that passes through a cell $C$ located two grid units away from $A$. In this configuration, the virtual-node mechanism becomes ineffective because the required referenced cell is beyond the face- or edge-adjacent neighborhood. Geometrically, this configuration requires the dihedral angle between the two sheets to be smaller than
$\omega=2\arctan{\frac{\sqrt{2}}{2}}\approx70.53^{\circ}$. We summarize the statistical distribution of dihedral angles for medial faces associated with internal feature lines, based on the results generated by MATStruct~\cite{wang2025matstruct}. Fig.~\ref{fig:virtual-node-angle-hist} shows the histogram of these angles, collected from 88 models and 52{,}173 angle samples. As shown in the histogram, dihedral angles below $70.53^\circ$ are extremely rare, accounting for only about $0.56\%$ of all samples. Moreover, even when a dihedral angle falls below $70.53^\circ$, it is very unlikely that exactly three non-empty neighboring nodes will align on the bisecting plane of the two sheets and along the direction perpendicular to their intersecting seam. Therefore, this specific failure scenario is extremely rare in practice, and we did not encounter any such cases in our experimental dataset.

\begin{table*}[h!]
\centering
\caption{Runtime comparison of different methods on selected ABC models (seconds). Our method uses an octree of maximum depth $7$, a default sample size  $n_\text{samples}=500{,}000$ of point cloud inputs, and KNN parameter $K=26$. Note that we display runtime of only MATTopo to generally demonstrate all RPD-based methods. Best values are shown in \textbf{bold}, second-best values are \underline{underlined}, and third-best values are shown in \textit{italics}.}
\label{tab:runtime_comparison}
\begin{tabular}{cccccccccc}
\toprule
Model   & MATTOPO & NSK    & PC\_160 & PC\_200 & SAT-0.03 & SAT-0.04 & VC    & VMAS & Ours  \\
\midrule
40051   & 174.13  & 182.49 & 32.07  & 43.99  & 10.93    & \textbf{6.37}     & \textit{8.47}  & 17.15 & \underline{7.74} \\
40052   & 56.25   & 107.77 & 7.12   & 10.22  & 6.63     & 4.25     & \textit{4.24}  & \textbf{2.94} & \underline{4.12} \\
40053   & 115.53  & 168.86 & 21.36  & 33.43  & 12.74    & \textit{7.30}     & \textbf{5.89}  & failed & \underline{7.00} \\
40057   & 119.75  & 155.51 & 17.41  & 26.38  & \textbf{0.35}     & \underline{5.27}     & \textit{5.98}  & 8.87 & 6.63 \\
40073   & 88.02   & 153.50 & 16.26  & 24.97  & \underline{6.62}     & \textbf{4.07}     & \textit{7.20}  & 33.33 & 8.46 \\
40085   & 98.54   & 135.85 & 18.41  & 26.95  & 11.05    & \textit{6.56}     & \underline{6.52}  & 6.65 & \textbf{5.94} \\
40092   & 265.50  & 69.25  & 34.67  & 50.14  & 13.50    & \textbf{8.12}     & \underline{8.97}  & 9.51 & 10.81 \\
40093   & 102.50  & 176.35 & 88.11  & 133.10 & 33.05    & \underline{19.83}    & \textbf{10.38} & failed & 21.17 \\
40097   & 141.01  & 204.73 & 39.57  & 57.59  & \textit{12.30}    & \textbf{7.05}     & 14.10 & 115.12 & \underline{11.38} \\
40108   & 123.75  & 81.11  & 22.31  & 33.49  & 10.48    & \underline{6.74}     & \textit{8.22}  & 122.47 & \textbf{5.18} \\
bear    & 52.50   & 82.53  & 18.10  & 26.81  & \textit{7.90}     & \underline{4.74}     & 9.80  & \textbf{1.87} & 13.2 \\
bug     & 75.00   & 158.70 & 13.79  & 21.01  & 5.73     & \underline{3.54}     & 8.67  & \textbf{1.00}  & \textit{4.49} \\
chair   & 62.42   & 148.91 & 16.88  & 26.52  & 6.24     & \textit{4.00}     & 8.96  & \textbf{1.64}  & \underline{3.54} \\
dolphin & 93.23   & 96.69  & 12.61  & 18.32  & \underline{5.99}     & \textbf{3.62}     & 8.39  & 15.32 & \textit{6.56} \\
hand    & 48.15   & 64.96  & 22.50  & 33.59  & \textit{7.22}     & \underline{3.91}     & 15.89 & \textbf{0.83} & 8.94  \\
plane   & 83.21   & 157.40 & 9.97   & 13.31  & 3.16     & \underline{1.88}     & 6.78  & \textbf{1.31} & \textit{3.01} \\
teapot  & 165.26  & 108.54 & 20.39  & 30.55  & \textit{10.96}    & \underline{6.59}     & 11.61 & \textbf{2.02} & 12.12 \\
venus   & 65.73   & 150.82 & 17.89  & 27.10  & \underline{9.19}     & \textbf{5.27}     & \textit{9.91}  & 11.29 & 10.97 \\
\bottomrule
\end{tabular}
\end{table*}

\subsection{Runtime Comparison}
To evaluate the efficiency of our method, we compare the medial axis extraction runtime on a selection of ABC models with existing baseline methods. Tab.~\ref{tab:runtime_comparison} reports the single-run processing time, in seconds, for several representative models. Our method is evaluated using the optimized C++ implementation with an octree of maximum depth $7$, a default input size of $n_\text{samples}=500{,}000$ oriented points, and KNN parameter $K=26$. All runtime experiments are conducted on Ubuntu to ensure consistent multithreading behavior and reproducible performance under OpenMP. We report MATTopo as a representative RPD-based method to indicate the typical runtime scale of this family of approaches.

As shown in Tab.~\ref{tab:runtime_comparison}, the optimized implementation of our method achieves competitive runtime across the tested models. Compared with MATTopo and NSK, our method is substantially faster on all listed examples. It is also generally faster than the Power Crust variants, especially on models where dense point samples lead to expensive reconstruction in PC\_160 and PC\_200. Compared with SAT, VC, and VMAS, our method is not uniformly the fastest, since these methods may use simpler or more aggressively simplified extraction procedures. However, our runtime remains in the same order of magnitude on most models while producing clean medial meshes with explicit sheet structure.

The remaining computational cost mainly comes from two stages. The first is the construction and processing of the KNN graph over approximately $500{,}000$ oriented samples with $K=26$, which is required for robust local grouping and feature-aware medial classification. The second is the solid-cell deletion stage, where each deletion requires updating edge--face adjacency tables and face-priority queues. These updates are partly sequential and therefore less amenable to parallelization. Nevertheless, after porting the major components to C++, the total runtime is reduced to a few seconds to about twenty seconds on most tested models, making the method practical for large-scale CAD inputs. 

\begin{figure}
\centering
\includegraphics[width=\linewidth]{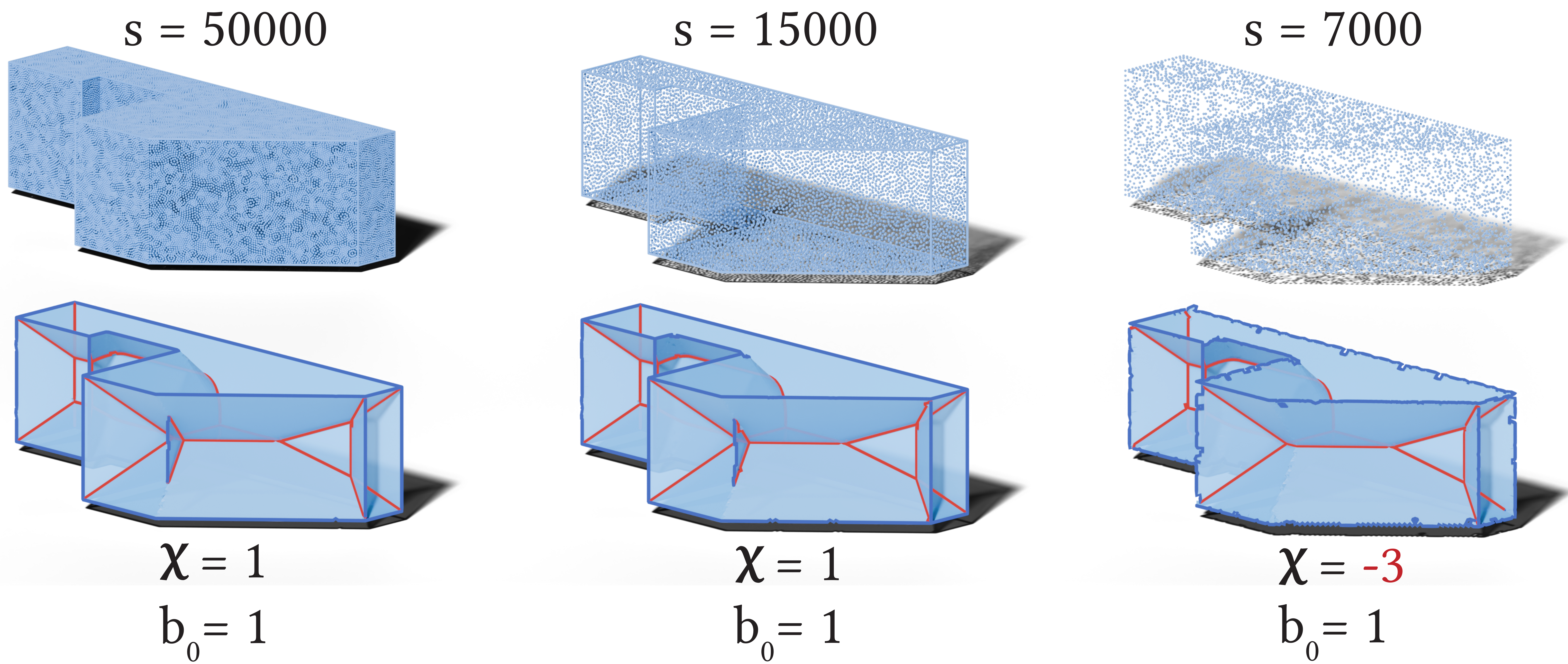}
\caption{\label{fig:sampling_convergence}Effect of point cloud's sampling density on medial axis reconstruction. Our computed medial meshes are shown for three different sample counts: 50,000, 15,000, and 7,000 points of input. Lower sampling densities produce incomplete sheets and minor local geometric inaccuracies; only when reduced to 7,000 points do topological errors begin to appear. Higher sampling densities stabilize the reconstruction and maintain sheet continuity and geometric fidelity.}
\end{figure}

\subsection{Limitations}
Although our method achieves high-quality medial axis reconstruction on oriented point clouds sampled from CAD-style or organic shapes, it still has several limitations:

\paragraph{Point Cloud Sampling}
We extract oriented point clouds from input meshes by detecting sharp features and uniformly sampling surface points. We perform a convergence analysis using $50{,}000$, $15{,}000$, and $7000$ points to investigate the influence of sampling density on medial mesh extraction. Fig.~\ref{fig:sampling_convergence} shows the resulting medial meshes: lower sampling densities produce incomplete sheets and locally inaccurate features, while higher densities stabilize the extraction.
\paragraph{Complex Tubular Regions and T$_\infty$ / 1D Medial Curve Connections} Our method may produce local breaks in connectivity when processing tubular regions. This occurs in both certain organic shapes and some CAD models. T$_\infty$ nodes are used to connect 1D medial curves. Both rely on heuristic strategies, which can occasionally cause incorrect connections or missed links. Note that the detection of T$_\infty$ nodes is not robust; the detection method is detailed in the Supplementary Document. Fig.~\ref{fig:T_infty_failure} illustrates a typical failure case: the red-marked medial segment is not properly connected via T$_\infty$, resulting in a local break of medial mesh.

\begin{figure}
\centering
\includegraphics{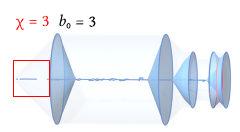}
\caption{\label{fig:T_infty_failure}Failure case in a complex tubular region involving $T_{\infty}$ / 1D medial curve completion.
The input shape contains elongated tubular structures where local $T_{\infty}$ detection and 1D edge connection become ambiguous.
The red-marked medial segment is not properly connected through the $T_{\infty}$-type completion step, resulting in a local break in the extracted medial mesh.}
\end{figure}

\paragraph{Parameter sensitivity} The method is sensitive to octree resolution and sampling density. With too few sample points, some critical nodes may contain only a single projection direction due to sparse sampling, causing them to be misclassified as empty. If the octree resolution is too low, medial structures with holes may fail to reconstruct (Fig.~\ref{fig:ablation_resolution}), affecting overall topological quality.

Overall, our method is suitable for oriented point clouds with clear features and sufficient sampling, but improvements are needed for complex tubular regions or noisy point clouds.

\begin{figure}
\centering
\includegraphics[width=\linewidth]{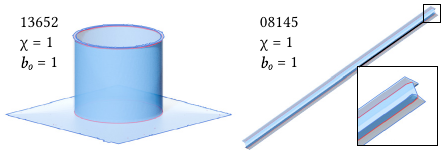}
\caption{\label{fig:mattopo_failures}
Challenging cases from the supplementary material of MATTopo~\cite{wang2024mattopo}. 
MATTopo fails to produce a valid medial axis on these inputs, while our method successfully extracts complete medial meshes directly from oriented point clouds without tetrahedral meshing or RPD computation. 
These examples illustrate the practical adaptability of DC-SQEM on inputs where volumetric RPD-based pipelines may fail.}
\end{figure} 

\subsection{More Results}
\begin{figure*}
    \centering
    \includegraphics[width=\textwidth]{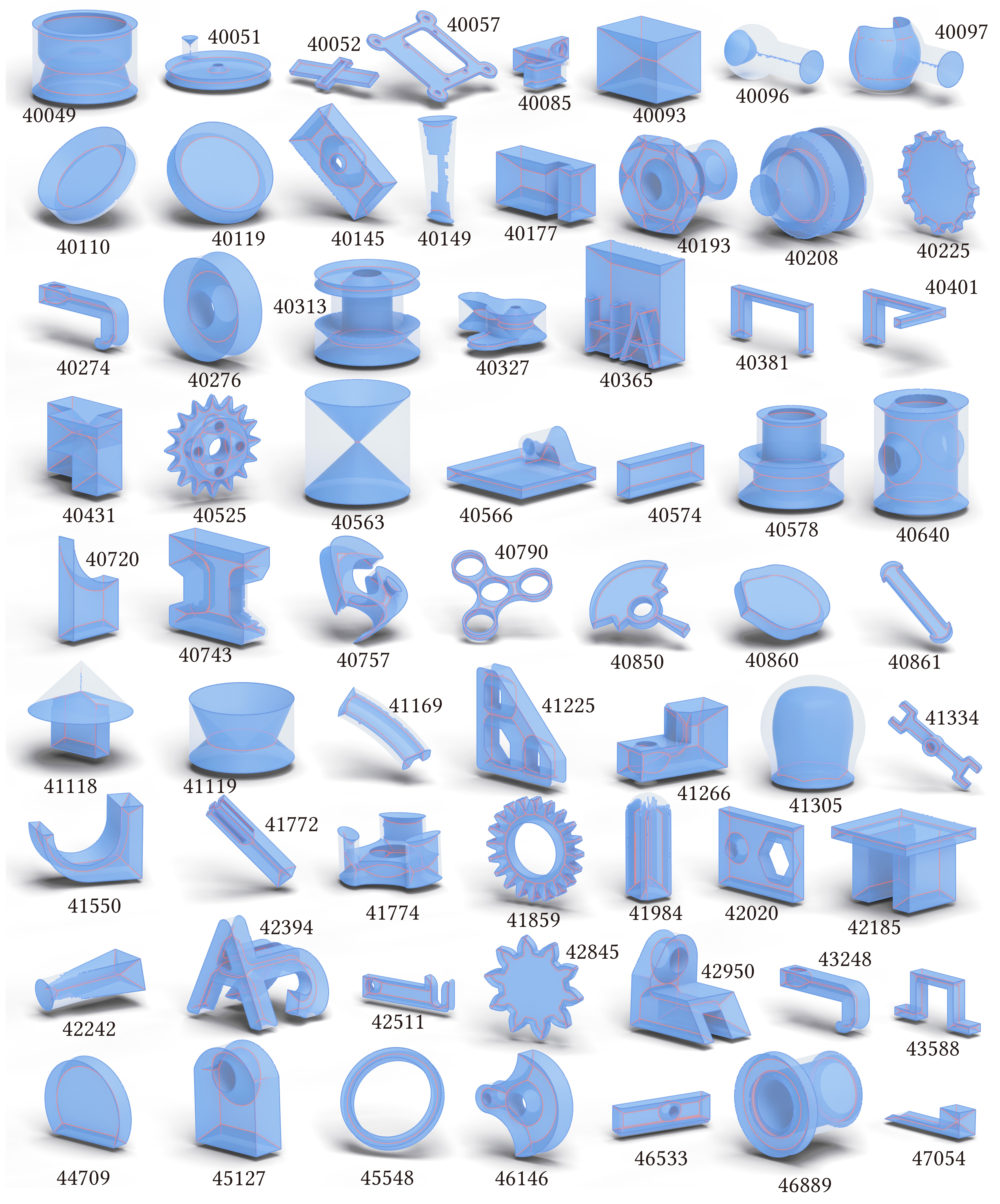}
    \caption{Gallery of medial meshes generated by our method on the remaining CAD models from the 93-model ABC benchmark. Models already shown on the teaser page and in the main text are excluded from this gallery.}
    \label{fig:abc gallery}
\end{figure*}

We provide more extensive visual results in Fig.~\ref{fig:abc gallery}. The gallery shows the medial meshes generated by our method on a diverse range of CAD shapes. We have renamed the original dataset identifiers to numeric IDs for clarity and brevity. The corresponding files for these results will be provided in the supplementary material for detailed inspection.
To further highlight the advantages of our method over MATTopo and MATStruct, we examine two challenging cases from the supplementary material of MATTopo~\cite{wang2024mattopo}, on which MATTopo fails to produce a valid medial axis. Methods such as MATTopo and MATStruct rely on tetrahedralization of input mesh and volumetric RPD computed on the tetrahedral mesh. Such RPD computations are numerically fragile when the shape is very thin. In contrast, our method only requires oriented point clouds and does not depend on tetrahedral meshing. As shown in Fig.~\ref{fig:mattopo_failures}, our method successfully generates complete and topologically sound medial meshes on these two inputs, demonstrating improved adaptability and scalability compared to MATTopo and MATStruct.

\section{Conclusion and Future Work}

We have presented \emph{DC-SQEM}, a framework for extracting medial meshes directly from oriented point clouds. By combining SQEM-based sphere fitting, virtual nodes, cell type downgrading, and topology- and geometry-aware thinning, DC-SQEM produces medial meshes that preserve sheets, curves, junctions, and seam structures with high geometric fidelity and low topological error rates.

Extensive evaluations on CAD and organic shapes show that DC-SQEM achieves consistently-high geometric accuracy (with low Hausdorff distance) and high triangle quality, while maintaining acceptable topological correctness relative to baseline methods, including MATFP, MATTopo, MATStruct, and VC. The method is robust to thin and curved regions, and consistently preserves medial structures of sheets, seams, and junctions.

There remain a few limitations to address, as connectivity in tubular regions with T$_\infty$ or 1D curves can be fragile, and our approach currently expects reasonably complete point cloud inputs. Future work will focus on designing a pipeline with rigorous topological guarantees for T$_\infty$ connectivity, further enhancing the method's robustness against heavily noisy or incomplete data.

\FloatBarrier

\begin{acks}
We thank anonymous reviewers for their detailed and constructive comments. This work was supported by the
\grantsponsor{fujian-major}{Fujian Provincial Science and Technology Major Project}{}
(Grant Nos.~\grantnum{fujian-major}{2024HZ022003} and
\grantnum{fujian-major}{2025HZ022030}),
the \grantsponsor{jiangxi-key}{Jiangxi Provincial Natural Science Foundation Key Project}{}
(Grant No.~\grantnum{jiangxi-key}{20244BAB28039}),
the \grantsponsor{xiamen-platform}{Xiamen Public Technology Service Platform}{}
(Grant No.~\grantnum{xiamen-platform}{3502Z20231043}),
and the \grantsponsor{fujian-sunshine}{Fujian Sunshine Charity Public Welfare Foundation}{}.
\end{acks}

\bibliographystyle{ACM-Reference-Format}
\bibliography{sample-base}

% \begin{table*}
%   \caption{Some Typical Commands}
%   \label{tab:commands}
%   \begin{tabular}{ccl}
%     \toprule
%     Command &A Number & Comments\\
%     \midrule
%     \texttt{{\char'134}author} & 100& Author \\
%     \texttt{{\char'134}table}& 300 & For tables\\
%     \texttt{{\char'134}table*}& 400& For wider tables\\
%     \bottomrule
%   \end{tabular}
% \end{table*}

%\bibliographystyle{ACM-Reference-Format}
%\bibliography{sample-base}

\appendix

\end{document}